\documentstyle[12pt,a41,epsfig,cite]{article}

\newcommand{\bea}{\begin{eqnarray}}
\newcommand{\bq}{\begin{equation}}
\newcommand{\eea}{\end{eqnarray}}
\newcommand{\eq}{\end{equation}}

\newcommand\GeV{\,\mbox{GeV}}
\newcommand\TeV{\,\mbox{TeV}}

\newcommand\MV{\mbox{\boldmath $M$}}

\newcommand\non{\nonumber}
\newcommand\del{\delta}
\newcommand\be{\begin{eqnarray}}
\newcommand\ee{\end{eqnarray}}
\newcommand\Sf{{\rm S}_{1,2}}
\newcommand\Sa{{\rm S}}

\newcommand\Li{{\rm Li}}
\newcommand\PV{\, \mbox{\boldmath $P$}}
\newcommand\DV{\, \mbox{\boldmath $D$}}
\newcommand\EV{\, \mbox{\boldmath $E$}}
\newcommand\UV{\, \mbox{\boldmath $U$}}
\newcommand\RV{\, \mbox{\boldmath $R$}}
\newcommand\M{\, \mbox{\boldmath $M$}}
\begin{document}
\noindent
\sloppy
\thispagestyle{empty}
\begin{flushleft}
DESY 02-016 \hfill
{\tt hep-ph/0409289}\\
SFB--CPP--04/11\\
January 2004
\end{flushleft}
%
\vspace*{\fill}
\begin{center}
{\LARGE\bf Universal Higher Order QED Corrections}

\vspace{2mm}
{\LARGE\bf to Polarized Lepton Scattering}

\vspace{2cm}
\large
Johannes Bl\"umlein  and
Hiroyuki Kawamura~\footnote{Present address~: KEK High Energy Accelerator 
Research Organization, 1--1 Oho, Tsukuba--shi, Ibaraki--ken 305--0901,
Japan.}
\\
\vspace{2em}
\normalsize
{\it Deutsches Elektronen--Synchrotron, DESY,\\
Platanenallee 6, D--15738 Zeuthen, Germany}
\\
\vspace{2em}
\end{center}
\vspace*{\fill}
%
\begin{abstract}
\noindent
We calculate the universal radiative QED corrections to polarized lepton 
scattering for general scattering cross sections in analytic form. The 
flavor non--singlet and singlet radiation functions are calculated to 
$O((\alpha {\rm L})^5)$. The resummation of the non--singlet and singlet 
contributions to the QED--anomalous dimensions $\propto(\alpha \ln^2(x))^k$ 
is performed to all orders. Numerical results are presented for the individual
radiation functions. Applications to polarized deeply inelastic lepton--nucleon 
scattering are given.
\end{abstract}
\vspace*{\fill}
\newpage
\section{Introduction}
\label{sec-1}

\vspace{1mm}
\noindent
QED corrections to integral and differential cross sections of light 
charged lepton--lepton scattering or deeply inelastic lepton--nucleon
scattering can be quite large in some kinematic
regions~\cite{EXAMP1,EXAMP2,RCJB}. 
In particular Bremsstrahlung contributions lead to significant shifts in the 
sub--system kinematics and cause large logarithmic contributions in 
differential distributions. The radiative corrections can be grouped into
universal pieces, which are process--independent, and the remainder
process dependent part. The former corrections are called universal
corrections as they are associated with the radiating particle 
irrespective of the process. Two classes of universal corrections are known~: 
{\it i)}~the leading--logarithmic $O\left[(\alpha L)^k\right]$ 
corrections, where $L = \ln(s/m^2)$. Here $s$ is a typical 
energy scale which describes a time-- or a space--like virtuality and $m$ 
denotes the light fermion mass.
This type of corrections can be obtained  by solving
the renormalization group equations for mass--factorization~\cite{CS} in
lowest order completely;
{\it ii)}~the QED--exponentiation  
$\left[\alpha \ln^2(x)\right]^k$--terms in the all--order anomalous dimensions
due to infrared evolution equations for initial--state radiation in lowest 
order~\cite{Kirschner:1983di,Bartels:1996wc}~\footnote{Numerical studies 
in the case of QCD were performed in~\cite{LNXQCD}. For a review see
\cite{Blumlein:1999ev}.}. 

Unlike the case in QCD, for QED radiation the radiation sources are no
extended 
distributions, $f_i(x,Q^2)$, which fall rapidly as $x \rightarrow 1$. The
sources are given by a $\delta(1-x)$--distribution instead. A numerical 
solution of the evolution equations via a Mellin transform with a complex 
argument $N$ and a subsequent numerical inverse Mellin transform is thus not 
possible since the latter requires substantial damping as ${\sf Re}(N) 
\rightarrow -\infty$, which is not the case in general here. To determine
the 
solution of the respective evolution equations in QED one rather evaluates
the 
required Mellin transforms of the evolution kernels analytically in
perturbative form up to an order in $(\alpha L)^k$ at which sufficient 
numerical stability is achieved. In the past the analytic resummation of the 
soft--
and virtual leading order contributions was 
performed~\cite{Gribov:1972rt}. In the
flavor non--singlet case analytic results were obtained to 5th order
in $\alpha L$~\cite{Przybycien:1993qe,Arbuzov:1999cq}~\footnote{
Earlier results up to 3rd order were given in 
\cite{Skrzypek:1992vk}.}. We compactify this 
description using the
small--$x$ resummations of terms $\propto \alpha \ln(x)$, as carried 
out in \cite{Jezabek:1992bx}
to 3rd order, to 5th order. However, these terms do not yield the 
complete description 
yet. The flavor singlet contributions have to be calculated 
to the same order, which is one of the objectives of the present paper.
  
The second class of universal corrections concerns the resummation of
contributions of the type $(\alpha \ln^2(x))^k$ in the cross section. These 
terms are resummed by infrared evolution equations. The flavor
non--singlet 
contributions were studied earlier in 
\cite{Blumlein:1996dd,Blumlein:1998yz}. 
As the comparison of
the first two terms of the resummation with QED 
corrections~\cite{Berends:1987ab} 
shows, the resummation makes a correct prediction for the terms which occur in
initial state radiation (ISR), whereas it remains an open problem, whether this is
the case for final state radiation (FSR)~\footnote{Note that these terms 
do potentially violate
the Gribov--Lipatov relation, see 
e.g.~\cite{Curci:1980uw,Blumlein:2000wh}.}. 
For this reason we
apply the resummation only for initial state radiation. In extension of 
earlier work in which the non--singlet resummation was dealt 
with~\cite{Blumlein:1998yz}
the resummation is carried out also for the singlet contributions.

The paper is organized as follows. In section~2 we revisit the non--singlet 
solution to $O\left[(\alpha L)^5\right]$ and compactify previous results
w.r.t. exponentiations for small and large values of $x$. Section~3
contains the 
general formalism to derive the singlet solution both for the 
leading order $O\left[(\alpha L)^k\right]$ terms and those due to the
resummation of contributions $\propto O\left[(\alpha \ln^2(x))^k\right]$  
consistently. The iterated singlet evolution kernels are presented
to $O\left[(\alpha L)^5\right]$ in section~4. In section~5 we derive the
coefficients for the resummation of the $O\left[(\alpha \ln^2(x))^k\right]$
terms for singlet evolution. In section~6 numerical results are presented
and section~7 contains the conclusions. Technical aspects of the calculation
being of interest also for other higher order QED and QCD  calculations
are summarized
in the Appendices where convolution tables, which were used in the present
calculation, are given in Appendix~A. As well we summarize a series of
higher order relations between Mellin transforms and harmonic sums, or
polynomials of them, which were used to calculate the more complicated
convolutions in Appendix B and C.

\section{The Non--Singlet Solution}

\vspace{1mm}\noindent
The universal radiative corrections discussed in the present paper can    
be derived using the renormalization group equations for mass             
factorization. We consider a one--flavor system out of electrons            
(positrons) and photons. The flavor non--singlet and singlet              
distributions  are defined by the following QED--distributions
\begin{eqnarray}            
\label{eqNSSD}                                                          
D_{\rm NS}^e(a,x) &=& D^{e^-}(a,x) - D^{e^+}(a,x) \\
D_{\Sigma}^e(a,x) &=& D^{e^-}(a,x) + D^{e^+}(a,x)~.
\end{eqnarray}
Here $x$ denotes the momentum fraction of the collinear
daughter--particle radiated off the source, $0 \leq x \leq 1$, 
and $a \equiv a(Q^2) = e^2/(4\pi)^2$ is the running coupling constant with
$e$ the electric charge. For this system only one non--singlet `minus'
distributions occurs.

The non--singlet contribution to the polarized QED evolution kernel
is the same as in the unpolarized case since the splitting functions
\begin{equation}
P^{ff}_{\rm NS}(x) = 2 \left(\frac{1+x^2}{1-x}\right)_+
\end{equation}
are identical. The $+$-operation is defined by
\begin{equation}
\int_0^1 dx \left[F(x)\right]_+ \phi(x) = \int_0^1 dx F(x)
\left[\phi(x)-\phi(1)\right]
\end{equation}
for the functions $\phi(x)$, $F(x)$ with $x~\epsilon~[0,1]$.

The leading order non--singlet evolution equation reads
\begin{equation}
\label{eqEVONS}
\frac{\partial}{\partial \ln(Q^2)} D_{\rm NS}^e(a,x) =
a(Q^2) P_{\rm NS}^{ff}(x) \otimes D_{\rm NS}^e(a,x)~.
\end{equation}
Here $\otimes$ denotes the Mellin--convolution
\begin{equation}
A(x) \otimes B(x) = \int_0^1 dx_1 \int_0^1 dx_2~
\delta(x-x_1 x_2) A(x_1) B(x_2)~.
\end{equation}
One may solve (\ref{eqEVONS}) using the Mellin--transform
\begin{equation}
\M[F(x)](N) = \int_0^1 dx x^{N-1} F(x)~.
\end{equation}
The Mellin--transform of the splitting function $P_{\rm NS}^{ff}(x)$ is
the anomalous dimension $\gamma_{\rm NS}^{ff}(N)$.

The QED running coupling constant is given by
\begin{equation}
\label{eqALP}   
\frac{d a}{d \ln(Q^2)} = - \sum_{k=0}^{\infty} \beta_k a^k
\end{equation}
with the first coefficients of the $\beta$--function $\beta_0 = -4/3,
\beta_1 
=-4$.
The solution of Eq.~(\ref{eqALP}) in leading order reads
\begin{equation}
\label{eqALPS}  
a(Q^2) = \frac{a_0}{\displaystyle 1- \frac{4}{3} a_0 L},
\end{equation}
with $a_0 \equiv a(m_e^2)$ and  $L = \ln(Q^2/m_e^2)$. For $a_0 L \ll 1$  
all later results obtained in $a$ may be expanded in the parameter
$a_0 L$ to a finite order.

In Mellin--space the solution of the non--singlet evolution equation
reads
\begin{equation}
\label{eqSOLNS}  
D_{\rm NS}^{ff}(N,Q^2) =\left[1 + \beta_0 a_0
\ln\left(\frac{Q^2}{m_e^2}\right)\right]^{\gamma_{\rm NS}^{ff}(N)/\beta_0}~.    
\end{equation}
Here we used the boundary condition $D_{\rm NS}^{ff}(N,m_e^2) = 1$ which
corresponds to
\begin{equation}
\label{eqINP}
D_{\rm NS}^{ff}(x,m_e^2) = \delta(1-x)~.
\end{equation}
For Eq.~(\ref{eqSOLNS}) one often defines $\hat{\beta} \equiv (2
\alpha_0/\pi)  
\ln(Q^2/m_e^2)$.

Unlike the case of QCD evolution of extended parton 
densities~\cite{Blumlein:1998em} or structure 
functions~\cite{Blumlein:2000wh},
which fall rapidly as $x \rightarrow 1$, the input distributions in QED
are of the form (\ref{eqINP}) and are distribution--valued
at $x=1$.
Therefore, the Mellin--inversion of (\ref{eqSOLNS}) cannot be performed
by a numerical contour integral but has to be worked out calculating all
the convolutions up to a finite order in $x$--space in to $(\alpha L)^k$ 
at 
which sufficient convergence of the series is being obtained. In the
present paper the iteration is performed up to $k=5$. The corresponding
convolutions are evaluated in explicit form in terms of Nielsen
integrals in Appendix~A. The calculation does partly require the
use of multiple harmonic 
sums~\cite{Vermaseren:1999uu,Blumlein:1999if,Blumlein:1997vf,ANCONT,ALGEBRA},
which can be
expressed as Mellin transforms of Nielsen integrals. In a previous
paper \cite{Blumlein:1999if} a series of these Mellin transforms has
been calculated already. In 
Appendix~B additional representations,
which were derived and used in the present calculation, are given.

We compared our results to those given previously in
\cite{Przybycien:1993qe,Arbuzov:1999cq}. We agree with 
\cite{Przybycien:1993qe} but find a deviation from \cite{Arbuzov:1999cq} in
the $O((\alpha L)^5)$ term. The result of \cite{Arbuzov:1999cq} is  
very lengthy and can be compactified by exponentiating
the soft-- and virtual contributions to all orders in $(\alpha L)^k$,
which was done before in Ref.~\cite{Przybycien:1993qe}. Moreover,
also the latter representation can be shortened in exponentiating also
the contributions $\propto (\hat{\beta} \ln(x))^l$, well--known from
QCD~\cite{DeRujula:1974rf}, which has been carried out for the first 
three orders in \cite{Jezabek:1992bx}, but is not contained in
\cite{Przybycien:1993qe,Arbuzov:1999cq}.

The non-singlet splitting kernel (no running coupling) can be represented
exponentiating the soft and small $x$ contributions to all orders and
treating the other contributions to finite order as
\begin{equation}
\label{eqdns}
D_{\rm NS}(x,\hat{\beta}) =
\left[\frac{\exp[(\hat{\beta}/2)(3/4-\gamma_E)]}
{\Gamma(1+\hat{\beta}/2)} \frac{\hat{\beta}}{2} (1-x)^{\hat{\beta}/2-1}
\frac{{\rm I}_1\left((-\hat{\beta}\ln(x))^{1/2}\right)}{[-\hat{\beta}
\ln(x)
]^{1/2}}
\sum_{n=0}^{\infty}    \left(\frac{\hat{\beta}}{2}\right)^n
\Psi_n(x)\right]_+
\end{equation}
with ${\rm I}_1(z)$ the associated Bessel function, $\gamma_E$ the
Euler--Mascheroni constant, and
\begin{eqnarray}
\Psi_0(x) &=& 1+x^2\\
\Psi_1(x) &=& -\frac{1}{2} \left[(1-x)^2+x^2\ln(x)\right]\\
\Psi_2(x) &=& \frac{1}{4} (1-x)\left[1-x-x\ln(x)+(1+x)\Li_2(1-x)\right]\\
\Psi_3(x) &=&  -\frac{1}{48} \Biggl\{6(1-x^2)[2 \Li_3(1-x)+ \Li_2(1-x)]
                          +5(x-1)^2 \\ & &
          + (1+7 x^2)[ \ln(x)\Li_2(1-x) + 2\Sf(1-x)]
          -\left(\frac{1}{2}+6 x - \frac{13}{2} x^2\right) \ln(x)
          +\frac{1}{12} x^2 \ln^3(x) \Biggr\} \nonumber\\
\Psi_4(x) &=&  \frac{1}{96} \Biggl\{(1-x^2)
\Biggl[24 \Li_4(1-x)+ 12\Li_3(1-x)
               - \frac{5}{2} \Sa_{1,3}(1-x) \nonumber\\ & &
               ~~~~~~~~~~~- 12 \Sa_{2,2}(1-x)
               - \frac{3}{2} \ln(x)\Sf(1-x)
              -\frac{1}{4} \ln^2(x) \Li_2(1-x)
              + 7 \Li_2(1-x)\Biggr] \nonumber  \\ & &
              ~~+ 4(1+x^2) \Li_2^2(1-x)
               +(1-8x+7x^2)\Biggl[\ln(x) \Li_2(1-x) +2\Sf(1-x)\Biggr]
               \nonumber\\ & &
              ~~+ 2 (1+7x^2)  \ln(x) \Li_3(1-x)
              -\left(\frac{3}{4} + 5x -\frac{23}{4} x^2\right) 
              \ln(x) \nonumber\\ & &
              ~~-\frac{1}{12}x(1-x) \ln^3(x)
              -\frac{1}{48}x^2 \ln^4(x)
              + (1-x)^2 \left[\frac{7}{2} 
              +\frac{1}{8}\ln^2(x) \right] \Biggr\}~.
\end{eqnarray}
Except the 0th term all the functions $\Psi_k(x)$ behave at least
as $\propto (1-x)$ for $x\rightarrow 1$. Therefore in these orders
the contribution $\propto 1/(1-x)$ in the function $D_{\rm
NS}(x,\hat{\beta})$ is
canceled. As mentioned before the leading small $x$ terms 
$\propto (\hat{\beta} \ln(x))^k$ are resummed into the function 
$I_1\left[(-\hat{\beta} \ln(x))^{1/2}\right]$.
Indeed many more logarithmic contributions are absorbed in this way
from the expanded expression. To 3rd order in $\hat{\beta}$ the functions
$\Psi_k(x)$ are finite as $x \rightarrow 0$. In the higher orders
some terms of the order $\hat{\beta}^k \ln^{k-l}(x), l > 0$ remain, but
contain a small weight factor only.
\section{The Solution of Singlet Evolution Equations}

\vspace{1mm}
\noindent
In this paragraph we derive the explicit solution of the singlet
evolution equation in terms of the running coupling constant $a=a(Q^2)$,
with $a= \alpha/(4\pi)$ and account both for the leading order terms
as well the leading terms at small $x$~~$\propto a^k \ln^{2(k-1)}(x)$.
Since the latter terms stem from the higher order evolution kernels,
which do not commute with the (resummed)
leading order solution in general, the perturbative solution has to be
constructed accordingly. A framework of this kind was derived for
Mellin $N$-space in Ref.~\cite{Blumlein:1998em} and for the lowest three
orders in \cite{Ellis:1994rb}. Here we will work in $x$--space and have
therefore to extend a prescription given to second order in  
\cite{Furmanski:1982cw} to all orders.

We substitute the scale dependence of the (singlet) evolution 
equation\footnote{The corresponding structures for the non--singlet
evolution equations are those of entry $(1,1)$ and are not given
separately. We will use this structures also resumming the small $x$ terms 
$\propto \ln^2(x)$ later on.}
in terms of $a$, Eq.~(\ref{eqALP}). One obtains
\begin{eqnarray}
\label{eqEVO1}
\frac{\partial \DV_S(a,x)}{\partial a} 
&=& 
- \frac{1}{a}
\frac{\PV_0(x) + a \PV_1(x) + a^2  \PV_2(x) + \ldots}
{\beta_0 + a \beta_1+ a^2 \beta_2 \ldots} \otimes \DV_S(a,x) \nonumber\\
&=& - \frac{1}{\beta_0 a} \Biggl[\PV_0(x) + a\left(\PV_1(x) 
- \frac{\beta_1}{\beta_0} \PV_0(x)\right) + O(a^2)\Biggr]
\otimes \DV_S(a,x) \nonumber\\
&=& - \frac{1}{a} \left[\RV_0(x) + \sum_{k=1}^{\infty} a^k \RV_k(x)\right]
\otimes \DV_S(a,x)~.
\end{eqnarray}
The functions $\RV_k(x)$ are given by
\begin{eqnarray}
\RV_0(x) &=& \frac{1}{\beta_0} \PV_0(x)\\
\RV_k(x) &=& \frac{1}{\beta_0} \PV_k(x) - \sum_{i=1}^k \frac{\beta_i}{
\beta_0} \RV_{k-i}(x)~.
\end{eqnarray}

The leading order solution is 
\begin{eqnarray}
\label{eqSLO}
\DV_{S,0}(a,x) &=& \left[\exp(-\RV_0(x) \ln(a/a_0)
)_\otimes\right] \otimes \DV_S(a_0,x) \equiv \EV_0(a,a_0,x) \otimes
\DV_S(a_0,x)~,
\end{eqnarray}
where $\EV_0$ is the leading order singlet evolution operator.
We use the short--hand notation
\begin{eqnarray}
\left[f(g(x))_\otimes\right] = \sum_{k=0}^{\infty} \frac{f^{(k)}(0)}{k!}
\otimes_{l=1}^k g(x)~,
\end{eqnarray}
with $\otimes_{l=1}^k$ the $k$-fold convolution. The distribution 
at the
scale $a = a_0$ is
\begin{eqnarray}
\DV_S(a_0,x) = {\bf 1}~\delta(1-x)~.
\end{eqnarray}

Higher than leading order contributions are found perturbativly by
\begin{eqnarray}
\DV_S(a,x) = \UV(a,x) \otimes \EV_0(a,a_0,x) \otimes
             \UV(a_0,x)^{-1}~,
\end{eqnarray}
with
\begin{eqnarray}
\label{eqUVk}
\UV(a,x) = {\bf 1}~\delta(1-x) + \sum_{k=1}^{\infty} a^k \UV_k(x)~.
\end{eqnarray}
The matrices $\UV_k(x)$ are obtained recursively from the functions
$\RV_k(x)$,
\begin{eqnarray}
\widehat{\RV}_1(x) &=& \RV_1(x) \\
\widehat{\RV}_k(x) &=& \RV_k(x) 
+ \sum_{l=1}^{k-1} \RV_{k-l} \otimes \UV_k(x)~.
\end{eqnarray}
In general the matrices $\UV_k$ do not commute. The higher than leading order
contributions dealt with in the present paper concern contributions to the
splitting functions $\propto a^k \ln^{2k}(x)$. If one would work in $x$--space
many convolutions of these functions with those representing the fixed order
solutions $\EV_0$ would have to be calculated. The series in $a^k \ln^{2k}(x)$ 
may need more terms than in the case of the fixed order solution in $O[(\alpha
L)^k]$. The specific structure of the terms $a^k \ln^{2k}(x)$ allows to work in
$N$--space since the Mellin transform
\begin{eqnarray}
\MV[\ln^{2k}(x)](N) = \frac{(2k)!}{N^{2k+1}}
\end{eqnarray}
is sufficiently damped for $k\geq 1$ in the limit ${\sf Re}(N) \rightarrow 
- \infty$. Due to this we may calculate the $\UV$--matrix in $N$--space and
use the solution
\begin{eqnarray}
\label{eqSOLII}
\DV_S(a,x) &=& \EV_0(a,a_0,x) +
\widehat{\UV}(a,x) \otimes \EV_0(a,a_0,x) +\EV_0(a,a_0,x) \otimes
             \widetilde{\UV}(a_0,x) \nonumber \\
& & 
+ \widehat{\UV}(a,x) \otimes \EV_0(a,a_0,x) \otimes \widetilde{\UV}(a_0,x)~,
\end{eqnarray}
where 
\begin{eqnarray}
\label{eqUV1}
\widehat{\UV}(a,x) &=& \UV(a,x) - {\bf 1} \delta(1-x) \\
\label{eqUV2}
\widetilde{\UV}(a,x) &=& \UV(a,x)^{-1} - {\bf 1} \delta(1-x)~.
\end{eqnarray}

\section{The Leading Order Solution to \mbox{\boldmath $O[(\alpha L)^5]$}}

\vspace{1mm}
\noindent
In the present paper we limit the explicit representations 
to order $n=5$ in $(a_0 L)$~~\footnote{Beyond this level we will use
only the exponentiation for the non--singlet terms.}. 
It is therefore convenient to expand (\ref{eqSLO}) as
\begin{eqnarray}
\label{EV0}
\EV_0(a,x) &=& {\bf 1}~
\delta(1-x) + \PV_0(x) a_0 L + \left( \frac{1}{2}
\PV_0^{(1)}(x) + \frac{2}{3} \PV_0(x)\right) (a_0 L)^2          \\ & &
+\left(\frac{1}{6} \PV_0^{(2)}(x)
      +\frac{2}{3} \PV_0^{(1)}(x)
      +\frac{16}{27} \PV_0(x)\right) (a_0 L)^3 \nonumber\\ & &
+\left(\frac{1}{24} \PV_0^{(3)}(x)
      +\frac{1}{3} \PV_0^{(2)}(x)
      +\frac{22}{27} \PV_0^{(1)}(x)
      +\frac{16}{27} \PV_0(x)\right) (a_0 L)^4 \nonumber\\ & &
+\left(\frac{1}{120} \PV_0^{(4)}(x)
      +\frac{1}{9} \PV_0^{(3)}(x)
      +\frac{14}{27} \PV_0^{(2)}(x)
      +\frac{80}{81} \PV_0^{(1)}(x)
      +\frac{256}{405} \PV_0(x)\right) (a_0 L)^5~. \nonumber
\end{eqnarray}
Eq.~(\ref{EV0}) makes the effect of the running coupling (\ref{eqALPS})
explicit. The matrices $\PV_0^{(k)}$ are
\begin{eqnarray}
\PV_0^{(k)}(x) = \otimes_{l=1}^k \PV_0(x)~.
\end{eqnarray}
The corresponding expressions for the $(1,1)$--components of
$\PV_0^{(k)}(x)$ are given relative to the non--singlet components
$\PV_{\rm NS}^{(k)}(x)$ of section~2. The singlet--matrix
components $\PV_{i,j}^{(k)}(x)$ given below~(\ref{eqSP0}--\ref{eqSP2}) 
were calculated using the convolution formulae of 
Appendix~A.~\footnote{Note that the second order contribution
$P_{1,1}^{(1)}(x)$ given in the literature is partly incorrect 
\cite{Akushevich:1998nz} for polarized scattering.}
The projections $P_{i,j}^{(k)}$ describe the splitting of an electron into 
an electron (1,1), of a photon into an electron, positron, respectively
(1,2), an 
electron into
a photon (2,1), and a photon into a photon (2,2) in $k$th order in the 
renormalized coupling constant. 

The individual components for the different orders are given by~:

\vspace{3mm} \noindent
{\sf 0th order~:}
\begin{eqnarray}
\label{eqSP0}
P_{1,1}^{(0)}(x) &=& P_{\rm NS}^{(0)}(x)  \\
P_{1,2}^{(0)}(x) &=& -4(1-2x)   \\
P_{2,1}^{(0)}(x) &=&  2(2-x)   \\
P_{2,2}^{(0)}(x) &=& - \frac{4}{3} \delta(1-x)~. 
\end{eqnarray}
Here a non--vanishing contribution to $P_{2,2}^{(0)}(x)$ emerges due to
4--momentum conservation, see also \cite{FADIN}.

\vspace{1mm} \noindent
{\sf 1st order~:}
\begin{eqnarray}
P_{1,1}^{(1)}(x) &=& P_{\rm NS}^{(1)} + 8[5(1-x)+2(1+x) \ln(x)]\\
P_{1,2}^{(1)}(x) &=& 4\left[2(2x-1)[2\ln(1-x)-\ln(x)] +\frac{13-8x}{3}
\right]\\
P_{2,1}^{(1)}(x) &=& 4\left[(2-x)[2\ln(1-x)+\ln(x)]-\frac{8-13x}{6}\right]\\
P_{2,2}^{(1)}(x) &=& 8\left[
            5(1-x)+2(1+x)\ln(x) + \frac{2}{9} \delta(1-x)\right]
\end{eqnarray}

\vspace{1mm} \noindent
{\sf 2nd order~:}
\begin{eqnarray}
P_{1,1}^{(2)}(x) &=& P_{\rm NS}^{(2)} 
+ 16\biggl[- \frac{49}{3}(1-x) 
+ 20(1-x)\ln(1-x) - \frac{4}{3}(7-8x)\ln(x) 
\\
&&\hspace{1cm} + 8(1+x)\ln(x)\ln(1-x) - 2(1+x)\ln^2(x) 
+ 8(1+x)\Li_2(1-x)\biggr] \non\\
P_{1,2}^{(2)}(x) &=& 8\left[\frac{1379-1318x}{18} + 8(1-2x)\zeta(2)  
+ \frac{44-16x}{3}\ln(1-x)+ \frac{98+158x}{3}\ln(x) \right.
\\
&& \left. - 8(1-2x)\ln^2(1-x) + 8(1-2x)\ln(x)\ln(1-x) 
+ 3(1-2x)\ln^2(x) \right]  \non\\
P_{2,1}^{(2)}(x) &=& 8 \left[ -\frac{659}{18} + \frac{1379}{36}x 
- 4(2-x)\zeta(2) - \frac{8-22x}{3}\ln(1-x) - \frac{71}{3}(1+x)\ln(x) \right.
\\
&&\left. + 4(2-x)\ln^2(1-x) - 4(2-x)\ln(x)\ln(1-x) 
- \frac{3}{2}(2-x)\ln^2(x) \right]   
\non\\
P_{2,2}^{(2)}(x) &=& 8\left[ -\frac{79}{3}(1-x) + 20(1-x)\ln(1-x) 
- \frac{20}{3}(2-x)\ln(x) \right.
\\ 
&& \left. + 8(1+x)\ln(x)\ln(1-x)
- 2(1+x)\ln^2(x) + 8(1+x)\Li_2(1-x) -\frac{8}{27}\delta(1-x) \right]~.  
\non
\end{eqnarray}

\vspace{1mm} \noindent
{\sf 3rd order~:}
\begin{eqnarray}
P_{1,1}^{(3)}(x) &=&  P_{\rm NS}^{(3)} 
+ 16\left[ -\left(\frac{5593}{18} + 120\zeta(2)\right)(1-x)
- \frac{548}{3}(1-x)\ln(1-x) \right.
\non\\                                
&& - \left(\frac{1445 + 3089x}{9} + 48(1+x)\zeta(2)\right)\ln(x)      
+ 120(1-x)\ln^2(1-x)                                               
\non\\                                                             
&& - \frac{80}{3}(4-5x)\ln(x)\ln(1-x) - \frac{85 - 65x}{3}\ln^2(x)  
- 24(1+x)\ln^2(x)\ln(1-x)                                          
\\                                                                
&& + 48(1+x)\ln(x)\ln^2(1-x) - \frac{2}{3}(1+x)\ln^3(x)         
+ \frac{40}{3}(1+x)\Li_2(1-x)                                 
\non\\                                                          
&& + 96(1+x)\ln(1-x)\Li_2(1-x)                            
- 96(1+x)\Li_3(1-x) + 48(1+x)\Sf(1-x) \biggl]  \non\\               
P_{1,2}^{(3)}(x) &=& 16\left[
-\frac{42587}{108}+\frac{10768}{27}x               
-\frac{124-32x}{3}\zeta(2)-32(1-2x)\zeta(3) \right.                           
\non\\                                                                
&& + \Biggl(\frac{2693 - 2506x}{9} + 48(1-2x)\zeta(2)\Biggr) \ln(1-x)   
\non\\
&& - \Biggl(\frac{3389 - 1384x}{18} + 24(1-2x)\zeta(2)\Biggr)\ln(x)
+ \frac{124-32x}{3}\ln^2(1 - x)
\\
&& - \frac{380 + 644x}{3}\ln(x)\ln(1 - x) 
-  \frac{217 + 304x}{6} \ln^2(x)
- 16(1-2x)\ln^3(1 - x) 
\non\\
&& + 24(1-2x)\ln(x)\ln^2(1 - x) 
+ 10(1-2x)\ln^2(x)\ln(1 - x) 
\non\\
&&
- \frac{7}{3}(1-2x)\ln^3(x)
+ (168+204x)\Li_2(1-x) + 44(1-2x)\ln(x)\Li_2(1-x)
\non\\
&&
+ 52(1-2x)\Sf(1-x) \biggr] \non\\
P_{2,1}^{(3)}(x) &=&  
16\left[
\frac{5384}{27}+\frac{42587}{216}x
+ \frac{16-62x}{3}\zeta(2)+16(2-x)\zeta(3) \right.
\non\\
&& - \Biggl(\frac{1253}{9} - \frac{2693}{18}x 
+ 24(2-x)\zeta(2)\Biggr)\ln(1-x)
\\
&&
+ \Biggr(\frac{907}{9}-\frac{1997}{36}x+12(2-x)\zeta(2)\Biggr)\ln(x)
- \frac{16 - 62x}{3}\ln^2(1-x)
\non\\
&&
- \frac{290 + 314x}{3}\ln(x)\ln(1 - x)
+ \Biggl(\frac{77}{3}+\frac{287}{12}x\Biggr)\ln^2(x)
+ 8(2-x)\ln^3(1 - x) 
\non\\
&&
- 12(2-x)\ln(x)\ln^2(1 - x)
- 5(2-x)\ln^2(x)\ln(1 - x)+\frac{7}{6}(2-x)\ln^3(x)
\non\\
&& - 22(2-x)\ln(x)\Li_2(1-x) - (102+84x)\Li_2(1-x)
- 26(2-x)\Sf(1-x)\biggr] \non\\
P_{2,2}^{(3)}(x) &=& 
16\left[ 
- \Biggl(\frac{2177}{6}+40\zeta(2)\Biggr)(1-x)
- \frac{236}{3}(1-x)\ln(1 - x) 
\right.\non\\
&&
- \Biggl(183+\frac{785}{3}x+16(1+x)\zeta(2)\Biggr)\ln(x)
+ 40(1-x)\ln^2(1 - x)
\\
&&
- \frac{128 - 112x}{3}\ln(x)\ln(1-x)
- \frac{103 - 107x}{3} \ln^2(x)
\non\\
&& +16(1+x)\ln(x)\ln^2(1 - x)
- 8(1+x)\ln^2(x)\ln(1 - x)-2(1+x)\ln^3(x)
\non\\
&& + 32(1+x)\ln(1 -x)\Li_2(1-x) 
-\frac{8}{3}(1+x)\Li2(1-x) - 32(1+x)\Li_3(1-x)
\non\\
&& \left. + 16(1+x)\Sf(1-x) + \frac{16}{81}\del(1-x) \right]~.    
\non
\end{eqnarray}

\vspace{1mm} \noindent
{\sf 4th order~:}
\begin{eqnarray}
P_{1,1}^{(4)}(x) &=&  P_{\rm NS}^{(4)} 
+ 32\left[\left(\frac{73678}{27}+704\zeta(2)+ 640\zeta(3)\right)(1-x)
\right.\non\\
&& - \left(\frac{15212}{9}+960\zeta(2)\right)(1-x)\ln(1 - x)
\non\\
&& + \left(\frac{36568}{27}- \frac{9068}{27}x 
+ 416\zeta(2)-544\zeta(2)x+ 256(1+x)\zeta(3)\right)\ln(x)
\non\\
&&
- 704(1-x)\ln^2(1 - x) 
- \left(\frac{8000+20672x}{9}+384(1+x)\zeta(2)\right)\ln(x)\ln(1-x)
\non\\
&&
+ \left(\frac{2696+4472x}{9} + 96(1+x)\right)\ln^2(x)
+ 320(1-x)\ln^3(1-x)
\non\\
&&
- (416-544x)\ln(x)\ln^2(1-x) - (152-88x) \ln^2(x)\ln(1-x)
\\
&& + \frac{40}{9}(8-7x)\ln^3(x)
+ 128(1+x)\ln(x)\ln^3(1 - x) - 96(1+x)\ln^2(x)\ln^2(1 - x)
\non\\
&& + \frac{4}{3}(1+x)\ln^4(x)
- \left(\frac{14336}{9}+384\zeta(2)\right)(1+x)\Li_2(1 - x) 
\non\\
&&
+ 128(1+x)\ln(1 - x)\Li_2(1 - x) - 720(1-x)\ln(x)\Li_2(1-x) 
\non\\
&& + 384(1+x)\ln^2(1-x)\Li_2(1-x) 
- 96(1+x)\ln^2(x)\Li_2(1-x)
\non\\
&& - 128(1+x)\Li_3(1 - x) -768(1+x)\ln(1-x)\Li_3(1-x) 
+ 768(1+x)\Li_4(1 - x) 
\non\\
&& - (816-944x)\Sf(1 - x) + 384(1+x)\ln(1-x)\Sf(1-x) 
\non\\
&&
-256(1+x)\ln(x)\Sf(1-x)
-384(1+x)\Sa_{2,2}(1 - x)-192(1+x)\Sa_{1,3}(1 - x)\Biggr]
\non\\
P_{1,2}^{(4)}(x) &=& 
32 \Biggl[
-\frac{1912189}{648} + \frac{958405}{324}x - \frac{7852-7064x}{9}\zeta(2) 
+ \frac{640-128x}{3}\zeta(3)- 48(1-2x)\zeta(4)
\non\\
&& + \left(-\frac{57280}{27}+\frac{58364}{27}x 
- 64(5-x)\zeta(2) - 256(1-2x)\zeta(3)\right)\ln(1-x) 
\non\\
&& + \left( -\frac{37633}{27}- \frac{100213}{27}x - 368\zeta(2) 
            - 656\zeta(2)x + 128(1-2x)\zeta(3) \right)\ln(x) 
\non\\
&& + \left( \frac{7852 -7064x}{9} + 192(1-2x)\zeta(2) \right)\ln^2(1-x)
\\
&& - \left( \frac{9220 - 4616x}{9} + 192(1-2x)\zeta(2) \right)\ln(x)\ln(1-x)
\non\\
&& + \left( -\frac{4703}{18}+ \frac{4442}{9}x 
                       - 24(1-2x)\zeta(2)\right)\ln^2(x)
+ \frac{64}{3}(5-x)\ln^3(1-x)
\non\\
&& + (368+656x)\ln(x)\ln^2(1-x) - \frac{616+1000x}{3}\ln^2(x)\ln(1-x)
\non\\
&& - \left(20+\frac{89}{3}x\right)\ln^3(x)
- 32(1-2x)\ln^4(1-x) + 64(1-2x)\ln(x)\ln^3(1-x)
\non\\
&& + 24(1-2x)\ln^2(x)\ln^2(1-x)
- \frac{40}{3}(1-2x)\ln^3(x)\ln(1-x) - \frac{5}{12}(1-2x)\ln^4(x)
\non\\
&& - (152+272x)\Li_2(1-x) + (1056+1248x)\ln(1-x)\Li_2(1-x)
\non\\
&& - \frac{32}{3}(4+x)\ln(x)\Li_2(1-x) + 288(1-2x)\ln(x)\ln(1-x)\Li_2(1-x) 
\non\\
&& - 16(1-2x)\ln^2(x)\Li_2(1-x) - (1056+1248x)\Li_3(1-x)
\non\\
&& - 288(1-2x)\ln(x)\Li_3(1-x) + 352(1-2x)\ln(1-x)\Sf(1-x)
\non\\
&&
+ 80(1-2x)\ln(x)\Sf(1-x) - 32(1-2x)\Li^2_2(1-x) 
+ 48(9 + 13x)\Sf(1-x)
\non\\
&& - 224(1-2x)\Sa_{2,2 }(1-x) + 224(1-2x)\Sa_{1,3}(1-x)\Biggr]
\non\\
P_{2,1}^{(4)}(x) &=& 
32 \Biggl[
\frac{958405}{648}-\frac{1912189}{1296}x+\frac{3532-3926x}{9}\zeta(2) 
- \frac{64}{3}(1-5x)\zeta(3)+24(2-x)\zeta(4) 
\non\\
&& + \left(\frac{41849+94913x}{54} + 296\zeta(2)+344\zeta(2)x 
                            -64(2-x)\zeta(3) \right)\ln(x)
\non\\
&& + \left(\frac{29182-28640x}{27}
                + 32(1-5x)\zeta(2) + 128(2-x)\zeta(3)\right) \ln(1-x)
\non\\
&& + \left(- \frac{3532-3926x}{9} - 96(2-x)\zeta(2) \right)\ln^2(1-x)
\\
&& + \left(\frac{4756-3242x}{9} + 96(2-x)\zeta(2)\right)\ln(x)\ln(1-x)
\non\\
&& + \left(\frac{1609}{9}-\frac{6071}{36}x +12(2-x)\zeta(2)\right)\ln^2(x)
- \frac{32}{3}(1 - 5x)\ln^3(1-x) 
\non\\
&& - (296+344x)\ln(x)\ln^2(1 - x) + \frac{436+484x}{3}\ln^2(x)\ln(1-x)
+ \left(\frac{35}{2}+\frac{14}{3}x\right)\ln^3(x)
\non\\
&& + 16(2-x)\ln^4(1- x) - 32(2-x)\ln(x)\ln^3(1 - x)
- 12(2-x)\ln^2(x)\ln^2(1-x)
\non\\
&& + \frac{20}{3}(2-x)\ln^3(x)\ln(1-x) + \frac{5}{24}(2-x)\ln^4(x)
+ (136+76x)\Li_2(1-x)
\non\\
&& - (624 + 528x)\ln(1-x)\Li_2(1-x) - \frac{16}{3}(1+4x)\ln(x)\Li_2(1-x) 
\non\\
&& - 144(2-x)\ln(x)\ln(1-x)\Li_2(1-x)
+ 8(2-x)\ln^2(x)\Li_2(1-x) 
\non\\
&&
+ 48(13 + 11x)\Li_3(1-x) + 144(2-x)\ln(x)\Li_3(1-x)
+ 16(2-x)\Li^2_2(1-x) 
\non\\
&& - 312(1+x)\Sf(1-x) - 176(2-x)\ln(1-x)\Sf(1-x) 
\non\\
&&
- 40(2-x)\ln(x)\Sf(1-x)
+ 112(2-x)\Sa_{2,2 }(1-x) - 112(2-x)\Sa_{1,3}(1-x) \Biggr]
\non\\
\label{eqSP2}
P_{2,2}^{(4)}(x) &=& 
32\Biggl[
\left(\frac{256273}{108}+ \frac{628}{3}\zeta(2)+160\zeta(3)\right)(1-x) 
-(1417+240\zeta(2))(1-x)\ln(1-x)
\non\\
&& + \left( \frac{32002-6257x}{27} + \frac{352-368x}{3}\zeta(2)
                               +64(1+x)\zeta(3) \right)\ln(x)
\non\\
&& - \frac{628}{3}(1-x)\ln^2(1 - x)
- \left(\frac{2150+3406x}{3}+96(1+x)\zeta(2)\right)\ln(x)\ln(1-x)
\non\\
&& + \left(\frac{763+626x}{3}+24(1+x)\zeta(2) \right)\ln^2(x)
+ 80(1-x)\ln^3(1-x) 
\\
&& - \frac{352-368x}{3}\ln(x)\ln^2(1-x) - \frac{394-386x}{3}\ln^2(x)\ln(1-x) 
\non\\
&& + \frac{50}{9}(5-4x)\ln^3(x) + 32(1+x)\ln(x)\ln^3(1-x)
\non\\
&& -24(1+x)\ln^2(x)\ln^2(1-x) - \frac{20}{3}(1+x)\ln^3(x)\ln(1-x)
+ \frac{7}{6}(1+x)\ln^4(x)
\non\\
&& - (926+96\zeta(2))(1+x)\Li_2(1-x) + \frac{16}{3}(1+x)\ln(1-x)\Li_2(1-x)
\non\\
&& - 380(1-x)\ln(x)\Li_2(1-x)
 + 96(1+x)\ln^2(1 - x)\Li_2(1-x)
\non\\
&& - 44(1+x)\ln^2(x)\Li_2(1-x) - \frac{16}{3}(1+x)\Li_3(1-x)
\non\\
&&
- 192(1+x)\ln(1-x)\Li_3(1-x) + 192(1+x)\Li_4(1-x) 
\non\\
&& - \frac{1252-1268x}{3}\Sf(1-x) + 96(1+x)\ln(1-x)\Sf(1-x)
\non\\
&& - 104(1+x)\ln(x)\Sf(1-x) -96(1+x)\Sa_{2,2 }(1-x)-88(1+x)\Sa_{1,3}(1-x)
\non\\
&&
-\frac{32}{243}\del(1-x)\Biggr]~.
\non
\end{eqnarray}
We expressed the above splitting functions in terms of poly--logarithms 
and Nielsen integrals~\cite{NIELS},
\begin{eqnarray}
\Li_n(x) &=& S_{n-1,1}(x) = \frac{(-1)^{n-1}}{(n-2)!} \int_0^1 
\frac{dz}{z} \ln^{n-2}
(z) \ln(1-xz) \\
S_{n,p}(x) &=& \frac{(-1)^{n+p-1}}{(n-1)! p!} \int_0^1\frac{dz}{z}
\ln^{n-1}(z) \ln^p(1-zx)~.
\end{eqnarray}
\section{\mbox{\boldmath $\alpha^{n+1}\ln^{2n}(x)$} Corrections to The
Evolution Kernel }

\vspace{1mm}
\noindent
The resummation of the leading small-$x$ terms to all orders in the
coupling constant were considered in QCD in
Refs.~\cite{LNXQCD,Blumlein:1996dd} and
\cite{Bartels:1996wc} for the singlet case. 
Very recently the results of \cite{Kirschner:1983di} in the interpretation 
of Ref.~[6a, 12] were verified to be correct in the 3rd order in the
coupling 
constant in the unpolarized non--singlet case \cite{OTHR}. This gives
further confidence in the method used.
The QED analogue is obtained by
setting $C_F = T_F =1, C_A =0$. We will furthermore consider only one
flavor, the electron.  
The corresponding contribution to the
matrix of singlet splitting functions reads
\begin{eqnarray}
{\bf P}(x,a)_{x \rightarrow 0} = \sum_{l=0}^{\infty}
{\bf P}^{(l)}_{x \rightarrow 0} a^{l+1} \ln^{2l}(x) = \frac{1}{8\pi^2}
{\cal M}^{-1}\left[F_0(N,a)\right](x).
\label{Eq1}
\end{eqnarray}

The matrix ${\bf F}_0(N,a)$ obeys the equation
\begin{eqnarray}
{\bf F}_0(N,a) = 16 \pi^2 \frac{a}{N} {\bf M}_0 - \frac{8a}{N^2}
{\bf F}_8(N,a){\bf G}_0 + \frac{1}{8\pi^2} \frac{1}{N} {\bf F}_0^2(N,a),
\label{Eq2}
\end{eqnarray}
where
\begin{eqnarray}
{\bf F}_8(N,a) = 16 \pi^2 \frac{a}{N} {\bf M}_8 
+ \frac{1}{8\pi^2} \frac{1}{N} {\bf F}_8^2(N,a).
\label{Eq3}
\end{eqnarray}
The generating  matrices are
\begin{eqnarray}
{\bf M}_0 = \left(\begin{array}{rr} 1 &  -2  \\ 2 & 0 \end{array}
\right),~~~~~~~~~~~{\bf M}_8 = \left( \begin{array}{rr} 1 & - 1 \\
0 & 0 \end{array} \right),~~~~~~~~~~~{\bf G}_0 = 
\left( \begin{array}{rr} 1 & 0 \\
0 & 0 \end{array} \right)
.
\label{Eq4}
\end{eqnarray}
Since ${\bf M}_8^n =  {\bf M}_8$ in the case of QED, ${\bf F}_8(N,a)$ 
can be given in analytic form
\be
\label{eqsqrt}
{\bf F}_8(N,a) = 4\pi^2 \left(1-\sqrt{1-\frac{8a}{N^2}}\right)
{\bf M}_8~.
\ee
\begin{center}
\renewcommand{\arraystretch}{1.5}
\begin{tabular}[h]{||r||r|r|r||}
\hline \hline
\multicolumn{1}{||c|}{$l$} &
\multicolumn{1}{c|}{$P_{ff}^{(l)}$} &
\multicolumn{1}{c|}{$P_{fg}^{(l)}$} &
\multicolumn{1}{c||}{$P_{gg}^{(l)}$} \\
\hline \hline
0  &
$ 0.2000000000$E$+01$  &
$-0.4000000000$E$+01$  &
$ 0.0000000000$E$+00$  \\
1  &
$-0.1400000000$E$+01$  &
$-0.4000000000$E$+01$  &
$-0.8000000000$E$+01$  \\
2  &
$-0.8666666667$E$+01$  &
$ 0.6666666667$E$+01$  &
$-0.2666666667$E$+01$  \\
3  &
$ 0.1444444444$E$+01$  &
$ 0.2444444444$E$+01$  &
$ 0.2044444444$E$+01$  \\
4  &
$ 0.7888888889$E$+00$  &
$-0.2825396825$E$+00$  &
$ 0.4634920635$E$+00$  \\
5  &
$ 0.5537918871$E$-02$  &
$-0.1008112875$E$+00$  &
$-0.3626102293$E$-01$  \\
6  &
$-0.1216503661$E$-01$  &
$-0.5536849981$E$-03$  &
$-0.9307893752$E$-02$  \\
7  &
$-0.5063924905$E$-03$  &
$ 0.8396483000$E$-03$  &
$-0.3947845218$E$-04$  \\
8  &
$ 0.4517396184$E$-04$  &
$ 0.2736857499$E$-04$  &
$ 0.4555056936$E$-04$  \\
9  &
$ 0.2611470148$E$-05$  &
$-9.1960462641$E$-05$  &
$ 0.1192943787$E$-05$  \\
10 &
$-0.4136771391$E$-07$  &
$-0.9306140011$E$-07$  &
$-0.6994805938$E$-07$  \\
11 &
$-0.4259181787$E$-08$  &
$ 0.1228221436$E$-08$  &
$-0.2774664782$E$-08$  \\
12 &
$-0.1716660622$E$-10$ &
$ 0.1073980741$E$-09$ &
$ 0.3096908050$E$-10$ \\
13 &
$ 0.2808782930$E$-11$ &
$ 0.3728781778$E$-12$ &
$ 0.2327857737$E$-11$ \\
14 &
$ 0.3713452748$E$-13$ &
$-0.5274589147$E$-13$ &
$ 0.7034915420$E$-14$ \\
\hline \hline
\end{tabular}
\end{center}

\vspace*{2mm}
\noindent
\begin{center}
{\sf Table~1: The coefficients of the matrices 
${\bf P}_{x \rightarrow 0}^{(l)}$ in Eq.~(\ref{Eq1}).}
\end{center}
\renewcommand{\arraystretch}{1}
Note that Eq.~(\ref{eqsqrt}) has no pole singularities but a branch cut
for
complex values of $N$. Due to this only moderate effects are expected.
The coefficient matrices  ${\bf P}^{(l)}_{x \rightarrow 0}$ are obtained
by iterative solution of Eqs.~(\ref{Eq2}).
The first coefficients read
\begin{eqnarray}
\label{Eq5}
{\bf P}^{(0)}_{x \rightarrow 0}
&=& 2 \left(\begin{array}{rr} 1 &  -2  \\ 2 & 0 \end{array}
\right) \\
\label{Eq6}
{\bf P}^{(1)}_{x \rightarrow 0}
&=& 2 \left(\begin{array}{rr} -4 &  -2  \\
2 & - 4  \end{array}
\right) \\
\label{Eq7}
{\bf P}^{(2)}_{x \rightarrow 0}
&=& \frac{2}{3}
  \left(\begin{array}{rr} -13 &  10  \\
-10 & -4  \end{array}
\right)~.
\end{eqnarray}
These matrices agree with the constant contribution to the leading
order polarized singlet splitting function and the coefficients of the
$\ln^2(x)$ contribution to the next-to-leading order splitting
function~\cite{QCD8} adjusting the color factors as above. As already
noted in the case of QCD the off-diagonal matrix elements are
related by
\begin{eqnarray}
\label{Eq8}
{\bf P}^{(l)}_{f\gamma} = - {\bf P}^{(l)}_{\gamma f}.
\end{eqnarray}
The first 15 coefficients are listed in Table~1.
The corrections $\propto a^{n+1} \ln^{2n}(x)$ are due to the most singular
parts of the non--leading order anomalous dimensions and are universal because
they emerge as part of a conformal solution. For each power in $a$ we have to
construct the corresponding matrices $\UV_k$,~Eq.~(\ref{eqUVk}). We use
Eqs.~(\ref{eqUV1}, \ref{eqUV2}) and work in $N$--space. The
Mellin--inversion is carried out by a numerical contour integral around
the singularities in the complex $N$--plane. Finally the solution is given by
Eq.~(\ref{eqSOLII}).

\section{Numerical Results}

\vspace{1mm}
\noindent
The calculation of the universal radiators is carried out to a precision
which will not need further improvement in the range of collider
energies at present facilities and those in the foreseeable future.
These corrections need to be supplemented by the non--universal
contributions of
the respective processes under consideration. In various cases the
universal contributions dominate in some kinematic regions. They do,
however, not substitute any complete calculation.    
In the above we presented the respective expressions in analytic form
which are easily implemented into QED radiators used in analysis
programs.  Now we turn to numerical illustrations.

The universal radiator function $D_{\rm NS}(x,Q^2)$,
Eq.~(\ref{eqdns}), summing the leading log corrections up to $O((\alpha
L)^5)$ is shown in Figure~1 as a function of the momentum
fraction $x$ for different values of $Q = \sqrt{Q^2} = 10, 100, 1000
\GeV$, the range of relevant scales at high energy colliders. The
function rises strongly for growing values of $x$ and reaches 1 for $x
\simeq 0.9$. $D_{\rm NS}(x,Q^2)$ grows with $Q$, however, the scale dependence 
is weaker. In Figure~2 we compare the soft photon resummation beyond the
5th order with $D_{\rm NS}(x,Q^2)$. Although these terms become more
significant for large values of $x$ the correction is widely below 3~ppm
in the range of $Q$ considered above, which shows that summing the leading
orders to $O((\alpha L)^5)$ for the full radiator is already a
sufficiently
accurate approximation.  In Figure~3 we show 
the relative impact of the radiator in first order compared to the
resummed radiator
for the same values of
$Q$. This ratio varies between $90$ to $95~\%$ for small
values of $x$ to $120$ to $130~\%$ for $x \rightarrow 1$. Also the the
4th
and 5th order approximation for $D^{k,\rm res}(x,Q^2)$ are compared,
including the respective soft exponentiation. The deviation is smaller
than
$1.5 \cdot 10^{-5}$ at small values of $x$ and vanishes for large values
of $x$, showing the degree of convergence.

In Figure~4 we show the singlet contributions for the leading logarithmic
radiators in case of  longitudinally polarized electrons or positrons.
Here, $D_{11}(x,Q^2)$ denotes the pure singlet part for the
fermion--radiator to which the non--singlet contribution has to be added.
The spread of the functions w.r.t. the scale variation in $Q$ is larger
for the off-diagonal radiators $D_{12}$ and $D_{21}$, as well as for
$D_{22}$ if compared to the fermionic radiator $D_{11}$. The off--diagonal
radiators $D_{12}(x)$ and $D_{21}(x)$ vary between $-6$ and $+6~\%$, and
$+7$ and $+2~\%$, respectively from low to large values of $x$. The pure
singlet part of the diagonal radiators is smaller. It grows from
values
of $-0.2~\% (-0.1~\%)$ for small values of $x$ to $+0.05~\%$ and 
vanish as $x \rightarrow 1$. The relative smallness of these contributions
results from the fact that they start with $O((\alpha L)^2)$ only, aside
of contributions $\propto \delta(1-x)$ not shown in the Figures.
 
Figure~5 compares the size of the first order contribution to $D_{11},
D_{12}$ and $D_{22}$ for the chosen values of $Q$ with the total
contribution. For $D_{11}$ the first order contribution is $20~\%$ larger
than the total contribution at small $x$ while it basically yields the
full contribution for large values of $x$. The first order
contribution to $D_{12}$ dominates $D_{12}$, except of the  small region
$x \sim 0.5$, where higher order contributions are significant. $D_{21}$
receives higher order corrections. $D_{21}^1/D_{21}^5 - 1$ varies from
$-5~\%$ to $+15~\%$ from small to large values of $x$. In Figure~6 we
compare the convergence of the approximation showing the impact of the 5th
order term w.r.t. the first four orders. This effect is of $O(10^{-5})$
with
some variations, where the effect in the case of $D_{22}$ is an order
stronger for large values of $x$. This shows that the this level of
resummation is sufficient.   

In Figure~7 we compare the radiators due to the resummation of the $\alpha
\ln^2(x) L$--terms for the same choice of scales $Q$ as before. 
$D_{11}(x,Q^2)$ denotes the complete fermionic 
radiator. The non--singlet contributions were dealt with before in
Ref.~\cite{Blumlein:1998yz}.
The pure singlet term contributes only with $O(\alpha^2)$.
Comparing Table~1 with 
Table~1~\cite{Blumlein:1998yz} one sees that the pure singlet contributions 
change the non--singlet result and is not subleading. We display the contributions starting
with $O(\alpha^2)$, i.e. those containing at least one term $\propto \ln^2(x)$. 
The first order contribution, the small $x$ limit of the leading order splitting functions,
has been contained in the radiators of the fixed order calculation already.
The respective part of the radiators $|D_{ij}(x,Q^2)|$ is of $O(1~\%)$ at
small values of $x$ and 
vanishes towards large values of $x$. 

We finally apply the corrections to polarized deeply inelastic
charged lepton scattering off polarized protons as an example. 
The hadronic tensor of the polarized part of the
$eN$--scattering cross section in case of pure photon
exchange\footnote{The corresponding expressions in the case of additional
weak boson exchange are given in \cite{BT}.} reads
\begin{eqnarray}
\label{eqHAD}
W^{(A)}_{\mu\nu} = i\varepsilon_{\mu\nu\lambda\sigma}
\frac{q^\lambda S^\sigma}{p.q} g_1(x,Q^2)
+i\varepsilon_{\mu\nu\lambda\sigma} \frac{q^\lambda (p.q
S^\sigma-S.q p^\sigma)}{(p.q)^2} g_2(x,Q^2)~.
\label{had1}
\end{eqnarray}
$S_\sigma$ denotes the nucleon spin vector, $p$ the nucleon momentum, and
$q$ the vector of the 4--momentum transfer, with $Q^2 = -q^2$ and
$x= Q^2/(2 p.q)$.
The polarized part of the scattering cross section for longitudinal
nucleon polarization $S_L$, integrated over the azimuthal angle $\phi$,
is
\begin{eqnarray} 
\label{eqdix}
\frac{d^2\sigma(\lambda, \pm S_L)_{NC}^B}{dxdy} &=& \pm 2
\pi{\sf S}
\frac{\alpha^2}{Q^4} \left [ -2\lambda y \left( 2-y-\frac{2 x y
M^2}{\sf S}
\right) xg_1(x,Q^2) + 8 \lambda \frac{y x^2 M^2}{\sf S} g_2(x,Q^2)
\right]~.\nonumber\\
\label{scaL} \end{eqnarray}
Here $M$ is the nucleon mass, $\sf S$ the cms energy, $\alpha$ the fine
structure constant, $y = 2p.q/{\sf S}$, $\lambda$ is the degree of lepton
lepton polarization, $x$ and $y$ are the Bjorken variables, 
and $g_{1,2}(x,Q^2)$ are the two 
electromagnetic polarized 
structure functions, which are represented at twist--2 level referring 
to the parameterization \cite{GRSV}. Very similar results are obtained
using other recent parameterizations \cite{AAC, BB}. We consider the case
of initial state 
radiation and sum over the final state bremsstrahlung. Furthermore, we cut
for the Compton peak, \cite{RCJB}. The kinematic variables chosen are  
the variables for leptonic corrections. 
The rescaled variables $\hat{x}, \hat{y}$, and $\hat{s}$ are
\begin{equation}
\hat{x} = \frac{xyz}{z+y-1},~~~~~~~~\hat{y} = \frac{z+y-1}{z},~~~~~~~\hat{s} = z{\sf S}~,
\end{equation}
and the radiation threshold is $z_0 = (1-y)/(1-xy)$~.
The radiative correction to the differential cross section (\ref{eqdix})
is given by~:
\begin{eqnarray}
\frac{d^2 \sigma_{NC}^{isr}}{dx dy} = \frac{d^2 \sigma_{NC}^{B}}{dx dy} 
+ \int_0^1 dz D(z,Q^2) \left\{\theta(z-z_0) {\cal J}(x,y,z) 
\left.\frac{d^2 \sigma_{NC}^{B}}{dx dy}\right|_{x=\hat{x},y=\hat{y},s=\hat{s}} -
\frac{d^2 \sigma_{NC}^{B}}{dx dy} \right\}~.
\end{eqnarray}
$D(z,Q^2)$ denotes the respective radiator function for fermion--fermion
transitions for 
$z < 1$, and the Jacobian ${\cal J}$ is given by
\begin{equation}
\label{eqjac}
{\cal J}(z,y,z) = \left | \begin{array}{cc}
\partial \hat{x}/ \partial x &
\partial \hat{y}/ \partial x \\
\partial \hat{x}/ \partial y &
\partial \hat{y}/ \partial y \end{array} \right |~.
\end{equation}
Previously,
the leptonic radiative corrections for this process have  been
calculated  in $O(\alpha)$ completely in Ref.~\cite{BBCK}, 
where the complete calculation was also compared to the leading
log result in $O(\alpha L)$. In the case of longitudinal polarization
studied in the present paper, both at HERMES energies,
$\sqrt{S} = 7.4 \GeV$, and for $ep$ collider
energies, $\sqrt{S} = 314~\GeV$, the complete $O(\alpha)$ corrections
deviate at most by a few~\% for small values of $x$ and very large values
of $y$ from the leading log result.

In Figure~8  the initial state QED corrections for the differential 
scattering cross section (\ref{eqdix}) are 
shown for the kinematics of the HERMES experiment at DESY with $\sqrt{S} = 
7.4 
\GeV$, 
as a ratio to the double differential Born cross section
using the resummed radiator as a function of $y$ for characteristic values 
of $x$. The radiative corrections grow with $y$ and towards smaller values 
of $x$ and may reach values of $O(80-100~\%)$ at large $y$. 
The effect of the higher than leading log corrections to those 
of leading order, normalized to the differential Born cross section is 
illustrated in the right figure.
Depending on the value of $x$, the corrections in the lower $y$ range are 
of up to $O(\pm 0.5~\%)$ and grow rapidly in the high $y$ region reaching 
$O(4~\%)$. 

For a hypothetical future $ep$ collider operating both with 
longitudinally polarized electrons and protons at a cms energy of $\sqrt{S} 
= 1~\TeV$ the double differential radiative corrections due to initial 
state 
radiation are shown in Figure~9. For fixed values of $x$ all the 
corrections grow strongly towards large value of $y$. For large values of 
$x$
they are negative for smaller values of $y$ and are everywhere positive 
for 
$x$ values below $x \sim 0.1$~. Due to the larger kinematic range the 
scaling violations due to the running fine--structure constant, 
$\alpha(Q^2)$,
are larger than in the case of the HERMES experiment. Even for $x$ values 
of $O(10^{-2})$ radiative corrections of $+160~\%$ may be reached for $y
\sim 0.9$.
We compare the impact of the higher than leading log corrections to those 
of leading order, normalized to the differential Born cross section.
Depending on the value of $x$, the corrections in the lower $y$ range are 
of up to $O(\pm 2.5~\%)$ and grow rapidly in the high $y$ region reaching 
$O(20~\%)$. 
\section{Conclusions}

\vspace{1mm}
\noindent
Mass factoriztion at the one hand and infrared evolution equations 
on the other hand, allow
to resum two different classes of {\sf universal} QED corrections
$\propto (\alpha \ln(s/m_e^2)^k$ and $\alpha^{k+1} \ln^{2k}(z)$, 
respectively, for characteristic invariant masses $s$ and radiation 
fractions 
$z$. The corresponding radiator functions are single--particle quantities 
and describe the universal part of the transitions between the light 
species, electrons $e$ and photons $\gamma$, via amplitudes
$D_{ij},~(i,j) = e, \gamma$, for collinear kinematics, similar to the 
parton 
model in QCD. In this way these radiators resum the universal part of the 
QED
parton cascades. We calculkated the non--singlet and singlet contributions 
for the case of polarized scattering to $O((\alpha L)^5)$ and resuummed the 
contributions $\propto \alpha^{k+1} \ln^{2k}(z)$. Both radiators are 
calculated to an accuracy which is sufficient for the foreseeable 
applications in present and future high energy experiments. While the 
calculation of the resummation of the $\alpha^{k+1} \ln^{2k}(z)$--terms can 
be done using Mellin space techniques, since the functions contributing are 
damped towards large Mellin parameters ${\sf Re}(N) \rightarrow - 
\infty$, this is not the case of for the radiator, resumming the
$(\alpha \ln(s/m_e^2)^k$--terms. Due to this the corresponding 
convolutions 
had to be calculated analytically in terms of Nielsen functions. The 
$\alpha^{k+1} \ln^{2k}(z)$--terms are leading small $z$ pieces of the higher 
order splitting functions. To deal with them in the context of the 
resummation of the $(\alpha \ln(s/m_e^2)^k$--terms, one has to invoke the 
mechanism of general higher order solutions of the singlet QED evolution 
equations for fermionic and photonic densites. Since the respective matrices 
of the splitting functions of the different orders in the coupling constant 
do not commute, it is practically excluded to write the solution in terms 
of analytic functions in $z$--space. However, one may solve the problem 
numerically and adds these contributions to the analytic radiators found for 
the $(\alpha \ln(s/m_e^2)^k$--terms. The main objective of the present 
paper 
was to provide all necessary radiators and to simplify results worked out 
earlier. The radiators obtained can easily be adopted for experimental 
analysis and 
simulation programs.

\newpage
\section{Appendix A:~Mellin Convolutions: Basic Functions}

\vspace{1mm}
\noindent
In this appendix we list the convolutions of functions up to depth~5
\footnote{A series of other useful convolutions has been given in
Ref.~\cite{Blumlein:2000wh} recently.},
which were calculated recursively in explicit form, since they are
of general interest for other higher order calculations in QED and QCD.
Some of the integrals
require to use Mellin transforms and algebraic relations between the
finite harmonic sums \cite{Blumlein:1999if},
being associated to them, to be solved. The respective expressions which
were used in this calculation are given in appendix~B and C, as long
they were not contained in Ref.~\cite{Blumlein:1999if}.
\begin{eqnarray}
f(x) \otimes \delta(1 -x ) &=& f(x),~~~\forall~
f(x)~\epsilon~{\cal S}'[0,1] \\
1 \otimes 1 &=& - \ln(x) \\
1 \otimes x &=&   1 - x \\
x \otimes x &=& - x \ln(x) \\
\left(\frac{1}{1-x}\right)_+ \otimes
\left(\frac{1}{1-x}\right)_+
&=& \Biggl[
2 \frac{\ln(1-x)}{1-x}
- \frac{\ln(x)}{1-x} \Biggr]_+
\\
\left(\frac{1}{1-x}\right)_+ \otimes 1 &=& \ln(1-x) - \ln(x) \\
\left(\frac{1}{1-x}\right)_+ \otimes x &=& x \left[\ln(1-x) - \ln(x) 
\right] + 1 - x \\
\left(\frac{\ln(1-x)}{1-x}\right)_+ \otimes  \left(\frac{1}{1-x}\right)_+
&=& \Biggl[
\frac{3}{2} \frac{\ln^2(1-x)}{1-x} - 
\frac{\zeta(2)}{1-x} 
- \ln(x) \frac{\ln(1-x)}{1-x} 
\Biggr]_+
\\
\left(\frac{\ln(1-x)}{1-x}\right)_+ \otimes  1
&=& \Li_2(x) - \zeta(2) +  \frac{1}{2} \ln^2(1-x)
\\
\left(\frac{\ln(1-x)}{1-x}\right)_+ \otimes  x
&=& x\Biggl[\Li_2(x) - \zeta(2) +  \frac{1}{2} \ln^2(1-x)-\ln(1-x)+\ln(x)\
\Biggr]\nonumber\\
& & \ \ +\ln(1-x)
\\
\left(\frac{1}{1-x}\right)_+ \otimes  \ln(1-x)
&=& \ln^2(1-x) - \ln(x) \ln(1-x) -     \zeta(2)
  \\
\left(\frac{1}{1-x}\right)_+ \otimes  x \ln(1-x)
&=&  x\Biggl[\ln^2(1-x) -\ln(x)\ln(1-x) - \zeta(2)\Biggr] 
\nonumber\\
& & + (1-x)[\ln(1-x)-1]
\\
\left(\frac{1}{1-x}\right)_+ \otimes  \ln(x)
&=& - \frac{1}{2} \ln^2(x) + \Li_2(1-x) + \ln(x) \ln(1-x)
\\
\left(\frac{1}{1-x}\right)_+ \otimes  x \ln(x)
&=& x \Biggl[ -\frac{1}{2} \ln^2(x) + \Li_2(1-x) + \ln(x) \ln(1-x) \Biggr]
\nonumber\\
& &-1+x-x\ln(x)
\\
\left(\frac{1}{1-x}\right)_+ \otimes  \frac{\ln(x)}{1-x}
&=& 
\ln(x) \frac{\ln(1-x)}{1-x} - \frac{1}{2} \frac{\ln^2(x)}{1-x}
  \\
1 \otimes \ln(1-x) &=& \Li_2(x) - \zeta(2)
 \\
1 \otimes x \ln(1-x) &=& (1-x) \left[\ln(1-x) - 1\right]
\\
x \otimes \ln(1-x) &=&  x \ln(x) + (1-x) \ln(1-x)
\\
x \otimes x \ln(1-x) &=&  x \left[\Li_2(x) - \zeta(2)\right]
\\
1 \otimes \frac{\ln(x)}{1-x} &=&  - \Li_2(1-x) - \frac{1}{2} \ln^2(x)
\end{eqnarray}
\begin{eqnarray}
x \otimes \frac{\ln(x)}{1-x} &=& 
x \Biggl[- \Li_2(1-x)-\frac{1}{2}\ln^2(x)\Biggr]+1 - x + \ln(x)
\\
1 \otimes \ln(x) &=&  - \frac{1}{2} \ln^2(x)
\\
1 \otimes x \ln(x) &=&  -1 + x - x \ln(x)
\\
x \otimes  \ln(x) &=& 1 - x + \ln(x) \\
x \otimes x \ln(x) &=&  - \frac{1}{2} x \ln^2(x)
\\
1 \otimes  \ln^2(x) &=&  - \frac{1}{3} \ln^3(x)
\\
x \otimes  \ln^2(x) &=& \ln^2(x) + 2 ( \ln(x) - x + 1)
\\
1 \otimes  x \ln^2(x) &=& - x \ln^2(x) + 2 (x \ln(x) - x +1)
\\
x \otimes  x \ln^2(x) &=&  - \frac{1}{3} x \ln^3(x)
\\
1 \otimes  \ln^3(x) &=&   - \frac{1}{4} \ln^4(x)
\\
x \otimes  \ln^3(x) &=&  \ln^3(x) + 3 \ln^2(x) + 6 (\ln(x) -x +1)
\\
1 \otimes  x \ln^3(x) &=&  - x \ln^3(x) + 3 x \ln^2(x) 
- 6(x \ln(x) - x +1)
\\
x \otimes  x \ln^3(x) &=&  - \frac{1}{4} x \ln^4(x)
\\
1 \otimes  \ln^4(x) &=&  - \frac{1}{5} \ln^5(x)
\\
x \otimes  \ln^4(x) &=&
 \ln^4(x) + 4 \ln^3(x) +12 \ln^2(x) +
24 (\ln(x) - x + 1)
\\
1 \otimes  x \ln^4(x) &=& 
- x \ln^4(x) + 4 x \ln^3(x) -12 x \ln^2(x) \nonumber\\ & &
+ 24 (x \ln(x) - x + 1)
\\
x \otimes  x \ln^4(x) &=& - \frac{1}{5} x \ln^5(x)
\\
\left(\frac{1}{1-x}\right)_+ \otimes \ln(x) \ln(1-x)
&=&  2\Sf(1-x) -\Li_3(1-x) \nonumber\\ & &
+ \left[\ln(x) + \ln(1-x)\right]
\Li_2(1-x) \nonumber\\  & &
-\frac{1}{2} \ln^2(x) \ln(1-x) + \ln(x) \ln^2(1-x) 
\nonumber\\ & &
- \zeta(2) \ln(x)
\\
\left(\frac{1}{1-x}\right)_+ \otimes x \ln(x) \ln(1-x)
&=&  x \Biggl\{2 \Sf(1-x) -\Li_3(1-x)\nonumber\\ & &
+ \left[\ln(x) + \ln(1-x)\right]
\Li_2(1-x) \nonumber\\  & &
-\frac{1}{2} \ln^2(x) \ln(1-x) + \ln(x) \ln^2(1-x) - \zeta(2) \ln(x)
\Biggr\} \nonumber\\
& &  + 2(1-x) - x \ln(x)\ln(1-x) + x\ln(x) \nonumber\\
& &  - (1-x) \ln(1-x) - \Li_2(1-x)
\\
\left(\frac{1}{1-x}\right)_+ \otimes \left(\frac{\ln^2(1-x)}{1-x}
\right)_+
&=&  \left[\frac{4}{3} \frac{\ln^3(1-x)}{1-x} - \ln(x) \frac{\ln^2(1-x)}
{1-x} \right.
\nonumber\\ & & \left. - 2 \zeta(2) \frac{\ln(1-x)}{1-x} + 2 \zeta(3)
\frac{1}{1-x}\right]_+
\\
\left(\frac{1}{1-x}\right)_+ \otimes \Li_2(1-x)
&=&  -\Sf(1-x) - \Li_3(1-x) \nonumber\\ & &
+\Li_2(1-x)[\ln(1-x) - \ln(x)]
\\
\left(\frac{1}{1-x}\right)_+ \otimes x \Li_2(1-x)
&=& -x\left[ \Sf(1-x) + \Li_3(1-x) \right.
\nonumber\\ & &  \left.
- \ln \left(\frac{1-x}{x}\right)
\Li_2(1-x)
\right] 
\nonumber\\ & &
-(1-x)[1-\Li_2(1-x)] - x \ln(x)
\\
\left(\frac{1}{1-x}\right)_+ \otimes \ln^2(1-x)
&=&  \ln^3(1-x)-\ln(x)\ln^2(1-x)
\nonumber\\ & &
-2\zeta(2)\ln(1-x)+2\zeta(3)
\\
\left(\frac{1}{1-x}\right)_+ \otimes x \ln^2(1-x)
&=& x \Biggl[ \ln^3(1-x)-\ln(x)\ln^2(1-x)
\nonumber\\&&
-2\zeta(2)\ln(1-x)+2\zeta(3)\Biggr]
\\ & &
+(1-x) \left[\ln^2(1-x) - 2 \ln(1-x) +2 \right]
\nonumber\\
\left(\frac{1}{1-x}\right)_+ \otimes \ln^2(x)
&=&  -2 \left[\Li_3(x) - \zeta(3) - \ln(x) \zeta(2) + \frac{1}{6}
\ln^3(x) \right]
\\
\left(\frac{1}{1-x}\right)_+ \otimes x \ln^2(x)
&=& -2x\Li_3(x) +2 \left[ \zeta(3)-1 \right] x  
+2\left[\zeta(2)+1\right]x\ln(x) \nonumber\\
& & -x\ln^2(x) - \frac{1}{3}x\ln^3(x)+2 
\\
\left(\frac{1}{1-x}\right)_+ \otimes \frac{\ln(x) \ln(1-x)}{1-x}
&=& 2\frac{\Sf(1-x)}{1-x}
+\frac{\ln(x)\Li_2(1-x)}{1-x} 
-\zeta(2)\frac{\ln(x)}{1-x}
\nonumber\\
& & -\frac{1}{2}\frac{\ln^2(x)\ln(1-x)}{1-x}+\frac{\ln(x)\ln^2(1-x)}{1-x}
\\
\left(\frac{1}{1-x}\right)_+ \otimes \frac{\ln^2(x)}{1-x}
&=& 2 \frac{\Sf(1-x)}{1-x}-\frac{1}{3}\frac{\ln^3(x)}{1-x}
\non\\
& & -2 \frac{\Li_3(x)-\zeta(3)}{1-x}+2\zeta(2)\frac{\ln(x)}{1-x}
\\
1 \otimes \ln(x) \ln(1-x)
&=&
\zeta(3) +   \Li_2(x) \ln(x) - \Li_3(x)
\\
1 \otimes x \ln(x) \ln(1-x)
&=&
-\Li_2(1-x) + (1-x)[2-\ln(1-x)] \nonumber\\ & &
+ x \ln(x) [1-\ln(1-x)]
\\
1 \otimes \left(\frac{\ln^2(1-x)}{1-x}\right)_+
&=&  - 2 \left[\Sf(x) - \zeta(3)\right] + \frac{1}{3} \ln^3(1-x)
\\
1 \otimes \Li_2(1-x)
&=&  -2\Sf(1-x) - \ln(x) \Li_2(1-x)
\\
1 \otimes x \Li_2(1-x)
&=&  (1-x) \Li_2(1-x) -1 + x - x\ln(x)
\\
1 \otimes \ln^2(1-x)
&=&  2 [ \zeta(3) - \Sf(x) ]
\\
1 \otimes x \ln^2(1-x)
&=& (1-x) \left[\ln^2(1-x) + 2  - 2 \ln(1-x)\right]
\\
1 \otimes \frac{\ln^2(x)}{1-x}
&=& 2\Sf(1-x)-\frac{1}{3}\ln^3(x)
\\
1 \otimes \frac{\ln(x) \ln(1-x)}{1-x}
&=& \Sf(1-x)-\Sf(x)\nonumber\\ 
& &+\frac{1}{2}\ln(x)\ln(1-x)\left[\ln(1-x)-\ln(x)\right]
+\zeta (3) 
\\
x \otimes \ln(x) \ln(1-x)
&=&  x \Li_2(1-x) + \frac{x}{2} \ln^2(x) + (1-x) \ln(1-x) \nonumber\\
& & + \ln(x)[x + \ln(1-x)]
\\
x \otimes x \ln(x) \ln(1-x)
&=&  x \left[\zeta(3) + \ln(x) \Li_2(x) - \Li_3(x) \right]
\\
x \otimes \frac{\ln^2(1-x)}{1-x}
&=&  2x [\zeta(3)-\Sf(x)] + (1-x) \ln^2(1-x) \nonumber\\ & &
+2x [\zeta(2) - \Li_2(x)] + \frac{x}{3} \ln^3(1-x)
\\
x \otimes \Li_2(1-x)
&=& (1-x) \Li_2(1-x) - \frac{1}{2} x \ln^2(x)
\\
x \otimes x \Li_2(1-x)
&=&  -x \left[2\Sf(1-x) + \ln(x) \Li_2(1-x)\right]
\\
x \otimes \ln^2(1-x)
&=&  
(1-x) \ln^2(1-x) - 2x \left[\Li_2(x)-\zeta(2)\right]
\\
x \otimes x \ln^2(1-x)
&=& 2x [ \zeta(3) - \Sf(x) ]
\\
x \otimes \frac{\ln(x) \ln(1-x)}{1-x}
&=& x\left\{\Sf(1-x)-\Sf(x) \right. \nonumber\\
& & \left. +\frac{1}{2}\ln(x)\ln(1-x)\left[\ln(1-x)-\ln(x)\right]
+\zeta (3) \right\} \\
& &+\ln(x)\ln(1-x)+\frac{x}{2}\ln^2(x)+(1-x)\ln(1-x)\nonumber\\
& &+x\ln(x)+x\Li_2(1-x) 
\nonumber\\
x \otimes \frac{\ln^2(x)}{1-x}
&=& x\left[2\Sf(1-x)-\frac{1}{3}\ln^3(x)\right]+\ln^2(x)\nonumber\\ & &
+2[\ln(x)+1-x]
\\
\left(\frac{1}{1-x}\right)_+ \otimes \left(\frac{\ln^3(1-x)}{1-x}\right)_+
&=&  \Biggl[\frac{5}{4} \frac{\ln^4(1-x)}{1-x} - \ln(x) \frac{\ln^3(1-x)}
{1-x} - 3 \zeta(2) \frac{\ln^2(1-x)}{1-x} \nonumber\\ & &
+6 \zeta(3) \frac{\ln(1-x)}{1-x} - 6 \zeta(4) \frac{1}{1-x} \Biggr]_+
\\
\left(\frac{1}{1-x}\right)_+ \otimes \frac{\ln(x)\ln^2(1-x)}{1-x}
&=& \frac{1}{1-x}
\Biggl[-2\Sa_{2,2}(x)-2\Sa_{2,2}(1-x)+\frac{\zeta(4)}{2}
\\ & & 
-\Li_2^2(1-x) + 6 \ln(1-x)\Sf(1-x)+ 4\ln(x)\Sf(x) 
\nonumber\\ & &
- 2\ln^2(x)\ln^2(1-x) + \ln(x)\ln^3(1-x) 
\nonumber\\ & &
- 2\zeta(2)\ln(x)\ln(1-x) \Biggr]
\nonumber\\
\left(\frac{1}{1-x}\right)_+ \otimes \frac{\ln^2(x)\ln(1-x)}{1-x}
&=& \frac{1}{1-x} 
\Biggl[ 4\ln(1-x)\Sf(1-x) -\Li_2(1-x)^2
\non\\ & &
+ 2\ln(x)[\Sf(1-x)+\Sf(x)-\zeta(3)]
\\ & &
- \ln^2(x)\Li_2(x) - \frac{4}{3}\ln^3(x)\ln(1-x) \Biggl]
\nonumber\\
\left(\frac{1}{1-x}\right)_+ \otimes \frac{\ln^3(x)}{1-x}
&=&  \frac{6}{1-x} \Biggl[-\Sa_{1,3}(1-x)+\zeta(4)-\Li_4(x) \nonumber\\
& & +\zeta(3) \ln(x) + \frac{1}{2} \zeta(2) \ln^2(x) - \frac{1}{24}
\ln^4(x) \Biggr]
\\
\left(\frac{1}{1-x}\right)_+ \otimes \frac{\Sf(1-x)}{1-x}
&=& \frac{1}{1-x}\Biggl[2 {\rm S}_{2,2}(1-x) - 2 {\rm S}_{1,3}(1-x)
\nonumber\\ & & - \frac{1}{2} \Li_2^2(1-x) + \ln\left(\frac{1-x}{x}\right)
{\rm S}_{1,2}(1-x) \Biggr]
\\
\left(\frac{1}{1-x}\right)_+ \otimes \frac{\Li_3(1-x)}{1-x}
&=& \frac{1}{1-x} \Biggl[\ln\left(\frac{1-x}{x}\right) \Li_3(1-x)
-{\rm S}_{2,2}(1-x)\Biggr]
\\
\left(\frac{1}{1-x}\right)_+ \otimes \frac{\ln(x)\Li_2(1-x)}{1-x}
&=& \frac{1}{1-x}\Biggl[2\Sa_{2,2}(x)-2\Sa_{2,2}(1-x)+4\Sa_{1,3}(1-x)
\nonumber\\ & &
+\frac{1}{2}\Li_2^2(1-x)
+\ln(x)[\Sf(1-x)-\Sf(x)-\zeta(3)]\nonumber\\ & &
-\frac{1}{2}\ln^2(x)\Li_2(1-x)
-2\ln(1-x)\Sf(1-x) \nonumber\\ & &
-\frac{1}{2}\zeta(4)\Biggr]
\\
\left(\frac{1}{1-x}\right)_+
\otimes \frac{\Li_3(x)-\zeta(3)}{1-x}
&=&\frac{1}{1-x}
\Biggl[
-2 \Sa_{2,2}(x) - 2 \Sa_{1,3}(1-x) 
\\ & &
- \ln(x) [\Sf(1-x) - 2\zeta(3)]
\nonumber\\ & &
- \ln(1-x)[\Sf(1-x) - \zeta(2)\ln(x)] 
+ \frac{1}{2} \zeta(4)
\nonumber \\ & &
+ \frac{1}{2}\Li_2(1-x)^2 + \frac{1}{6}\ln^3(x)\ln(1-x)
\Biggr]
\nonumber\\
\left(\frac{1}{1-x}\right)_+ \otimes \Sf(x)
&=&  
\Sa_{2,2}(x) - 3 \Sa_{1,3}(x) - \ln(x) \Sf(x) \nonumber\\ & &
+\zeta(2) \Li_2(x) + \zeta(3) \ln(1-x) + \frac{1}{4}\zeta(4)
\\
\left(\frac{1}{1-x}\right)_+ \otimes x \Sf(x)
&=&  x\Biggl\{
\Sa_{2,2}(x) - 3 \Sa_{1,3}(x) - \ln(x) \Sf(x) \nonumber\\ & &
\zeta(2) \Li_2(x) + \zeta(3) \ln(1-x) + \frac{1}{4}\zeta(4) \Biggr\}
\nonumber\\ & &
+\zeta(3) - x \Sf(x) \nonumber\\ & &
- (1-x)\left[1-\ln(1-x) + \frac{1}{2} \ln^2(1-x)\right]
\\
\left(\frac{1}{1-x}\right)_+ \otimes \Li_3(1-x)
&=& -{\rm S}_{2,2}(1-x) - \Li_4(1-x) + \ln\left(\frac{1-x}{x}\right)
\Li_3(1-x)
\\
\left(\frac{1}{1-x}\right)_+ \otimes x \Li_3(1-x)
&=& x \Biggl\{
-{\rm S}_{2,2}(1-x) - \Li_4(1-x) \nonumber\\ & &
+ \ln\left(\frac{1-x}{x}\right)
\Li_3(1-x) \Biggr\} \nonumber\\
& & + x\ln(x) + (1-x)[1-\Li_2(1-x)+\Li_3(1-x)]
\\
\left(\frac{1}{1-x}\right)_+ \otimes \ln(1-x)\Li_2(1-x)
&=&  
2\Sa_{2,2}(x)-\Sa_{1,3}(x) +\frac{\zeta(4)}{2}
\\ &&
- 3\ln(1-x)\Sf(1-x) -2\ln(x)\Sf(x) +\frac{1}{2}\Li_2(1-x)^2
\nonumber\\ & &  
+\left[\frac{1}{2}\ln^2(1-x)-\ln(x)\ln(1-x)-\zeta(2)\right]\Li_2(1-x)
\non\\ & &  
-\frac{1}{6}\ln(x)\ln^3(1-x)+\frac{1}{2}\ln^2(x)\ln^2(1-x)
\nonumber\\
\left(\frac{1}{1-x}\right)_+ \otimes x \ln(1-x)\Li_2(1-x)
&=&x\Biggl\{
2\Sa_{2,2}(x)-\Sa_{1,3}(x) +\frac{\zeta(4)}{2}
\\ &&
- 3\ln(1-x)\Sf(1-x) -2\ln(x)\Sf(x) +\frac{1}{2}\Li_2(1-x)^2
\nonumber\\ & &  
+\left[\frac{1}{2}\ln^2(1-x)-\ln(x)\ln(1-x)-\zeta(2)\right]\Li_2(1-x)
\non\\ & &  
-\frac{1}{6}\ln(x)\ln^3(1-x)+\frac{1}{2}\ln^2(x)\ln^2(1-x)\Biggr\}
\non\\ & &
+(1-x)\left[\ln(1-x)-1\right]\Li_2(1-x)-\Li_2(1-x)
\non\\ & &
+3(1-x)+2x\ln(x)-[1-x+x\ln(x)]\ln(1-x)
\nonumber\\
\left(\frac{1}{1-x}\right)_+ \otimes \ln(x)\Li_2(1-x)
&=& 2\Sa_{2,2}(x)-2\Sa_{2,2}(1-x)+4\Sa_{1,3}(1-x)+\Li_2^2(1-x)
\nonumber\\&&
+\ln(x)[\Sf(1-x)-\Sf(x)-\zeta(3)]\nonumber\\ & &
-\frac{1}{2}\ln^2(x)\Li_2(1-x)
\nonumber \\&&
-2\ln(1-x)\Sf(1-x)-\frac{1}{2}\zeta(4)
\\
\left(\frac{1}{1-x}\right)_+ \otimes x \ln(x)\Li_2(1-x) 
&=& x\Biggl\{2\Sa_{2,2}(x)-2\Sa_{2,2}(1-x)+4\Sa_{1,3}(1-x)
\nonumber\\&&
+\ln(x)[\Sf(1-x)-\Sf(x)-\zeta(3)] \nonumber\\ & &
+\Li_2^2(1-x)
-\frac{1}{2}\ln^2(x)\Li_2(1-x)
\nonumber \\&&
-2\ln(1-x)\Sf(1-x)-\frac{1}{2}\zeta(4)\Biggr\}
\\&&
+[1-x+x\ln(x)](3-\Li_2(1-x))-x\ln^2(x) \nonumber\\ & &
-2\Sf(1-x)
\nonumber\\
\left(\frac{1}{1-x}\right)_+ \otimes \Sf(1-x)
&=& -2 {\rm S}_{1,3}(1-x) + {\rm S}_{2,2}(1-x) 
- \frac{1}{2} \Li_2^2(1-x) \nonumber\\ & &
+ \ln\left(\frac{1-x}{x}\right) \Sf(1-x)
\\
\left(\frac{1}{1-x}\right)_+ \otimes x \Sf(1-x)
&=&  x \Biggl[
-2 {\rm S}_{1,3}(1-x) + {\rm S}_{2,2}(1-x) 
- \frac{1}{2} \Li_2^2(1-x) \nonumber\\ & &
+ \ln\left(\frac{1-x}{x}\right) \Sf(1-x) \Biggr] \nonumber\\ & &
+ (1-x)[\Sf(1-x)-1] - x \ln(x) \left[1 - \frac{1}{2} \ln(x)\right]
\\
\left(\frac{1}{1-x}\right)_+ \otimes \Li_3(x)
&=& 
-2 \Sa_{2,2}(x) - 2 \Sa_{1,3}(1-x) - \ln(x) \Sf(1-x)
\non\\ & &
- \zeta(2)\Li_2(x) + \frac{1}{6} \ln^3(x)\ln(1-x) + 3\zeta(4) 
\\ & &
+ \zeta(3)[\ln(x)+\ln(1-x)]
\non\\
\left(\frac{1}{1-x}\right)_+ \otimes x \Li_3(x)
&=& x\Biggl[
-2 \Sa_{2,2}(x) - 2 \Sa_{1,3}(1-x) - \ln(x) \Sf(1-x)
\non\\ & &
- \zeta(2) \Li_2(x) + \frac{1}{6} \ln^3(x)\ln(1-x) + 3 \zeta(4)
\\ && 
+ \zeta(3)[\ln(x) + \ln(1-x)] 
\Biggr]  
\non\\ & &
+\zeta(3)-\zeta(2)+ x[\Li_2(x)-\Li_3(x)]
\non\\ & &
+(1-x)[1-\ln(1-x)]
\non\\
\left(\frac{1}{1-x}\right)_+ \otimes \ln(x)\Li_2(x)
&=& -2\Sa_{2,2}(x)-4\Sa_{1,3}(1-x)
\non\\&&
+\ln(x)[3\zeta(3)-\Sf(x)-3\Sf(1-x)]
\\ & &
+ \frac{1}{2} \ln^2(x)\Li_2(x) + 3\zeta(4) - \zeta(2)\Li_2(x)
\non\\ &&
+ \frac{5}{6}\ln^3(x)\ln(1-x)
\nonumber\\
\left(\frac{1}{1-x}\right)_+ \otimes x \ln(x)\Li_2(x)
&=& x\Biggl\{
-2\Sa_{2,2}(x)-4\Sa_{1,3}(1-x)
\non\\&&
+\ln(x)[3\zeta(3)-\Sf(x)-3\Sf(1-x)]
\nonumber\\ & &
+ \frac{1}{2}\ln^2(x)\Li_2(x)
+3\zeta(4)-\zeta(2)\Li_2(x) 
\\ & &
+ \frac{5}{6}\ln^3(x)\ln(1-x)
\Biggr\}
+[1+x-x\ln(x)]\Li_2(x)
\non\\&&
+(1-x)(2-\ln(x))[1-\ln(1-x)]
\nonumber\\ &&
-2\zeta(2)+1-x+\ln(x)
\nonumber\\
\left(\frac{1}{1-x}\right)_+ \otimes \ln^3(1-x)
&=&  \ln^4(1-x) - \ln(x) \ln^3(1-x) - 3\zeta(2) \ln^2(1-x) \nonumber\\
& & + 6\zeta(3) \ln(1-x) - 6 \zeta(4)
\\
\left(\frac{1}{1-x}\right)_+ \otimes x \ln^3(1-x)
&=&  x \Biggl[
\ln^4(1-x) - \ln(x) \ln^3(1-x) - 3\zeta(2) \ln^2(1-x) \nonumber\\
& & + 6\zeta(3) \ln(1-x) - 6 \zeta(4) \Biggr ]
\nonumber\\ & &
- 6(1-x)\Biggl[1 - \ln(1-x) + \frac{1}{2} \ln^2(1-x) \nonumber\\
& & - \frac{1}{6} \ln^3(1-x) \Biggr]
\\
\left(\frac{1}{1-x}\right)_+ \otimes \ln(x)\ln^2(1-x)
&=& 
-2\Sa_{2,2}(x)-2\Sa_{2,2}(1-x)-2\Sa_{1,3}(x)+\frac{5}{2}\zeta(2)
\non\\&&
-\Li_2^2(1-x) + 6\ln(1-x)\Sf(1-x)
\\&& 
- 4 \ln(x)[\Li_3(1-x)-\zeta(3)]
\non\\&&
+2\ln(x)\ln(1-x)\left[2\Li_2(1-x)-\zeta(2)\right]
\nonumber\\ & & 
+\frac{2}{3}\ln(x)\ln^3(1-x)
\nonumber\\
\left(\frac{1}{1-x}\right)_+ \otimes x \ln(x)\ln^2(1-x)
&=&x\Biggl\{
-2\Sa_{2,2}(x)-2\Sa_{2,2}(1-x)-2\Sa_{1,3}(x)+\frac{5}{2}\zeta(2)
\non\\&&
-\Li_2^2(1-x) + 6\ln(1-x)\Sf(1-x)
\\&& 
- 4 \ln(x)[\Li_3(1-x)-\zeta(3)]
\non\\&&
+2\ln(x)\ln(1-x)\left[2\Li_2(1-x)-\zeta(2)\right]
\nonumber\\ & & 
+\frac{2}{3}\ln(x)\ln^3(1-x)
\Biggr\}-2x\ln(x)
\nonumber\\ & & 
+(1-x)\ln(x)\ln(1-x)[\ln(1-x)-2]
\nonumber\\ & & 
-(1-x)[\ln(1-x)^2-4\ln(1-x)+6]
\nonumber\\& & 
+2[\zeta(2)-\Li_2(x)]+2[\zeta(3)-\Sf(x)]
\nonumber\\ 
\left(\frac{1}{1-x}\right)_+ \otimes \ln^2(x)\ln(1-x)
&=& 
2\Sa_{2,2}(1-x)+2[\ln(x)+\ln(1-x)]\Sf(1-x)
\\&&
+ 2\ln(x)[\Sf(x)-\zeta(3)]-\Li_2(1-x)^2
\non\\&&
+\ln^2(x)[\Li_2(1-x)-\zeta(2)]
-\frac{1}{3}\ln^3(x)\ln(1-x)
\nonumber\\
\left(\frac{1}{1-x}\right)_+ \otimes x \ln^2(x)\ln(1-x)
&=& x\Biggl\{
2\Sa_{2,2}(1-x)+2[\ln(x)+\ln(1-x)]\Sf(1-x)
\non\\&&
+ 2\ln(x)[\Sf(x)-\zeta(3)]-\Li_2(1-x)^2
\\&&
+\ln^2(x)[\Li_2(1-x)-\zeta(2)]
-\frac{1}{3}\ln^3(x)\ln(1-x)
\Biggr\}
\nonumber\\ & &
+2\Sf(1-x)+2\Li_2(1-x)
\non\\&&
-x[\ln^2(x)-2\ln(x)]\ln(1-x)+2(1-x)\ln(1-x)
\non\\&&
-4x\ln(x)+x\ln^2(x)+6(x-1)
\nonumber\\
\left(\frac{1}{1-x}\right)_+ \otimes \ln^3(x)
&=& 
6 \Biggl[
\zeta(4) - \Li_4(x) + \zeta(3) \ln(x) + \frac{1}{2} \zeta(2)
\ln^2(x) \Biggr]
 \nonumber\\ & & 
   - \frac{1}{4} \ln^4(x)
\\
\left(\frac{1}{1-x}\right)_+ \otimes x \ln^3(x)
&=&
6x \Biggl[
\zeta(4) - \Li_4(x) + \zeta(3) \ln(x) + \frac{1}{2} \zeta(2)
\ln^2(x)
 \nonumber\\ & & + 1 - \ln(x) + \frac{1}{2} \ln^2(x) - \frac{1}{6}
 \ln^3(x) \Biggr] \nonumber\\ & &
-6 - \frac{x}{4} \ln^4(x)
\\
1 \otimes \left(\frac{\ln^3(1-x)}{1-x}\right)_+
&=& 6 \left[{\rm S}_{1,3}(x) - \zeta(4) \right] + \frac{1}{4} \ln^4(1-x)
\\
1 \otimes \frac{\ln(x)\ln^2(1-x)}{1-x}
&=& 2 \Biggl[
{\Sa}_{2,2}(x) +\Sa_{1,3}(x) -\ln(x) \Sf(x)
-\frac{5}{4}\zeta(4) 
\non \\ & &
+\frac{1}{6}\ln(x)\ln^3(1-x)\Biggr]
\\
1 \otimes \frac{\ln^2(x)\ln(1-x)}{1-x}
&=& 
- 2 \Biggl[ \Sa_{2,2}(1-x) + \Sa_{1,3}(1-x) 
\\ & &
- \ln(1-x) \Sf(1-x)
+ \frac{1}{6}\ln^3(x)\ln(1-x)\Biggr]
\non\\
1 \otimes \frac{\ln^3(x)}{1-x}
&=& - \frac{1}{4} \ln^4(x) - 6 {\rm S}_{1,3}(1-x)
\\
1 \otimes \frac{\Sf(1-x)}{1-x}
&=& {\rm S}_{2,2}(1-x) - \ln(x) \Sf(1-x) - 3 {\rm S}_{1,3}(1-x)
\\
1 \otimes \frac{\Li_3(1-x)}{1-x}
&=& - \frac{1}{2} \Li_2^2(1-x) - \ln(x) \Li_3(1-x) + \Li_4(1-x)
\\
1 \otimes \frac{\ln(x)\Li_2(1-x)}{1-x}
&=& - \frac{1}{2} \Li_2^2(1-x) - \frac{1}{2} \ln^2(x) \Li_2(1-x)
+ 3 {\rm S}_{1,3}(1-x)
\\
1 \otimes \frac{\Li_3(x)-\zeta(3)}{1-x}
&=& \zeta(4) - \Li_4(x) + \zeta(3) \ln(x) + \ln(1-x) [\Li_3(x) - \zeta(3)
] \nonumber\\ & & -\frac{1}{2} [\zeta(2)^2 - \Li_2^2(x)]
\\
1 \otimes \Sf(x)
&=& \frac{1}{4} \zeta(4) - {\rm S}_{2,2}(x)
\\
1 \otimes x \Sf(x)
&=&   \zeta(3) - x\Sf(x)  \nonumber\\ & &
- (1-x) \left[ 1 - \ln(1-x)
+ \frac{1}{2} \ln^2(1-x) \right]
\\
1 \otimes \Li_3(1-x)
&=& - \ln(x) \Li_3(1-x)  - \frac{1}{2} \Li_2^2(1-x)
\\
1 \otimes x \Li_3(1-x)
&=& x \ln(x) +(1-x)[1 - \Li_2(1-x) + \Li_3(1-x)]
\\
1 \otimes \ln(1-x)\Li_2(1-x)
&=& 
2\Sa_{2,2}(1-x) - \frac{1}{2}\Li^2_2(1-x) - 2\ln(1-x) \Sf(1-x)
\nonumber\\ & &
-\ln(x)\ln(1-x) \Li_2(1-x)
\\
1 \otimes x \ln(1-x)\Li_2(1-x)
&=& - \Li_2(1-x) + (1-x)[\ln(1-x) - 1] \Li_2(1-x) \nonumber\\ & &
+ (1-x)[3-\ln(1-x)]
+ x \ln(x) [2 - \ln(1-x)]
\\
1 \otimes \ln(x)\Li_2(1-x)
&=& -\frac{1}{2} \ln^2(x) \Li_2(1-x) + 3 {\rm S}_{1,3}(1-x)
\\
1 \otimes x \ln(x)\Li_2(1-x) 
&=& 3(1-x) + x \ln(x)[3-\ln(x)] \nonumber\\ & &
- x[\ln(x)-1] \Li_2(1-x)
                -2 \Sf(1-x) - \Li_2(1-x)  \nonumber\\
\\
1 \otimes \Sf(1-x)
&=& - \ln(x) \Sf(1-x) - 3 \Sa_{1,3}(1-x)
\\
1 \otimes x \Sf(1-x)
&=& (1-x)[\Sf(1-x) - 1) 
- x\left[\ln(x) - \frac{1}{2} \ln^2(x)  \right]
\\
1 \otimes \Li_3(x)
&=& \zeta(4)-\Li_4(x)
\\
1 \otimes x \Li_3(x)
&=& (1-x)[1-\ln(1-x)] + x [\Li_2(x)- \Li_3(x)] \nonumber\\
& & +\zeta(3) - \zeta(2)
\\
1 \otimes \ln(x)\Li_2(x)
&=& \Li_4(x) - \ln(x) \Li_3(x) - \zeta(4)
\\
1 \otimes x \ln(x)\Li_2(x)
&=& [3 - 2\ln(1-x)](1-x) - \Li_2(1-x) - \zeta(2) \nonumber\\ & &
+x \left\{\Li_2(x) \left[1 - \ln(x) \right] + \ln(x) \left[1-
\ln(1-x)\right] \right\}
\\
1 \otimes \ln^3(1-x)
&=&  -6 [\zeta(4) - {\rm S}_{1,3}(x)]
\\
1 \otimes x \ln^3(1-x)
&=& -(1-x)[6 - 6 \ln(1-x) + 3 \ln^2(1-x) \nonumber\\ & &
- \ln^3(1-x)]
\\
1 \otimes \ln(x)\ln^2(1-x)
&=& -2\left[\frac{1}{4} \zeta(4) -{\rm S}_{2,2}(x) + \ln(x) \Sf(x)\right]
\\
1 \otimes x \ln(x)\ln^2(1-x)
&=& 2[\zeta(3) - \Sf(x)] + 2[\zeta(2)-\Li_2(x)] - 2\ln(x) \nonumber\\ & &
-(1-x)[6 - 4 \ln(1-x) + \ln^2(1-x)] \nonumber\\ & &
+ (1-x) \ln(x) [2 - 2 \ln(1-x) + \ln^2(1-x)]
\\
1 \otimes \ln^2(x)\ln(1-x)
&=& - 2 [ \Sa_{1,3}(1-x)+\frac{1}{6}\ln^3(x)\ln(1-x)]
\\
1 \otimes x \ln^2(x)\ln(1-x)
&=& 2 \Biggl[\Sf(1-x) + \Li_2(1-x)\Biggr]
\nonumber\\ & & +x \ln(x) \Biggl\{-2 + \biggl[2-\ln(x)\biggr]\biggl[
\ln(1-x)-1\biggr]\Biggr\} 
\\ & &
+2(1-x)\left[\ln(1-x)-3\right]
\non\\
x \otimes \left(\frac{\ln^3(1-x)}{1-x}\right)_+
&=& 6x \left[\Sa_{1,3}(x) + \Sa_{1,2}(x) - \zeta(4)  -\zeta(3) \right]
\nonumber\\ & &
+ \frac{x}{4} \ln^4(1-x) + (1-x) \ln^3(1-x)
\\
x \otimes \frac{\ln(x)\ln^2(1-x)}{1-x}
&=& 2 x \Biggl[
\Sa_{2,2}(x) + \Sa_{1,3}(x) -\ln(x) \Sf(x) - \frac{5}{4}\zeta(4) 
\non\\ & &  
- \Sf(1-x) + \Sf(x) + \frac{1}{6}\ln(x)\ln^3(1-x)
\\& &
-\frac{1}{2}\ln(x)\ln(1-x)\left[\ln(1-x)-\ln(x)\right]
-\zeta (3)\Biggr]
\nonumber\\ & &  
+(1-x) \ln^2(1-x) - 2x \left[\Li_2(x)-\zeta(2)\right] \nonumber\\
& &+ \ln(x) \ln^2(1-x)
\non\\
x \otimes \frac{\ln^2(x)\ln(1-x)}{1-x}
&=& -2 x \Biggl[
\Sa_{2,2}(1-x) + \Sa_{1,3}(1-x) - \ln(1-x) \Sf(1-x)
\nonumber\\ &&
+\Sf(1-x)-\Li_2(1-x)+\frac{1}{6}\ln^3(x)\ln(1-x)
\\&&
-\frac{1}{6}\ln^3(x)-\frac{1}{2}\ln^2(x)+\ln(1-x)-\ln(x)
\Biggr]
\non\\&&
+2[1+\ln(x)]\ln(1-x)+\ln^2(x)\ln(1-x)
\non\\
x \otimes \frac{\ln^3(x)}{1-x}
&=& - x \left[\frac{1}{4} \ln^4(x) + 6 {\rm S}_{1,3}(1-x) \right]
\nonumber\\ & &
+6 \left[1-x+\ln(x) + \frac{1}{2} \ln^2(x) + \frac{1}{6} \ln^3(x)
\right]
\\
x \otimes \frac{\Sf(1-x)}{1-x}
&=& x \Biggl[{\rm S}_{2,2}(1-x) - \ln(x) \Sf(1-x) - 3 {\rm S}_{1,3}(1-x)
\Biggr] \nonumber\\ & & + \left({1-x} \right) \Sf(1-x)
+ \frac{x}{6} \ln^3(x)
\\
x \otimes \frac{\Li_3(1-x)}{1-x}
&=& - x \left[\frac{1}{2} \Li_2^2(1-x) + \ln(x) \Li_3(1-x) - \Li_4(1-x)
\right] \nonumber\\ & &
+(1-x) \Li_3(1-x) + x \ln(x) \Li_2(1-x) \nonumber\\ & &
+ 2x \Sf(1-x) \nonumber\\
\\
x \otimes \frac{\ln(x)\Li_2(1-x)}{1-x}
&=& x \Biggl[-\frac{1}{2} \Li_2^2(1-x) 
- \frac{1}{2} \ln^2(x) \Li_2(1-x) + 3 {\rm S}_{1,3}(1-x) \Biggr]
\nonumber\\ & & +[1 + \ln(x) ] \Li_2(1-x) - \frac{x}{3} \ln^3(x)
- \frac{x}{2} \ln^2(x) \nonumber\\ & &
+2x \Sf(1-x) - x \Li_2(1-x)
\\
x \otimes \frac{\Li_3(x)-\zeta(3)}{1-x}
&=& \Li_3(x) + \Li_2(x) - \zeta(3) -x \zeta(2) - x\ln(x) \nonumber\\ & &
-(1-x) \ln(1-x) \nonumber\\  & &
+ x \Biggl\{
\zeta(4) - \Li_4(x) + \zeta(3) \ln(x) \nonumber\\ & &
+ \ln(1-x) [\Li_3(x) - \zeta(3)
] \nonumber\\ & & -\frac{1}{2} [\zeta(2)^2 - \Li_2^2(x)]\Biggr\}
\\
x \otimes \Sf(x)
&=& {\rm S}_{1,2}(x) - x \zeta(3) + \frac{1}{2}(1-x) \ln^2(1-x) +
x[\zeta(2) \nonumber\\ & &
- \Li_2(x)]
\\
x \otimes x \Sf(x)
&=& x\left[\frac{1}{4} \zeta(4) - {\rm S}_{2,2}(x) \right]
\\
x \otimes \Li_3(1-x)
&=& 2x \Sf(1-x) +(1-x) \Li_3(1-x) \nonumber\\ & &
+x \ln(x) \Li_2(1-x)
\\
x \otimes x \Li_3(1-x)
&=& - x\left[\ln(x) \Li_3(1-x)  + \frac{1}{2} \Li_2^2(1-x)\right] 
\\
x \otimes \ln(1-x)\Li_2(1-x)
&=& 3x \Sf(1-x) 
- \frac{x}{2} \ln^2(x) \ln(1-x)
\nonumber\\ & &
+ \left[x\ln(x) + (1-x) \ln(1-x) \right] \Li_2(1-x)
\\
x \otimes x \ln(1-x)\Li_2(1-x)
&=& x \Biggl[
2\Sa_{2,2}(1-x) - \frac{1}{2}\Li_2^2(1-x) \nonumber\\ & &
- 2\ln(1-x) \Sf(1-x)
\nonumber\\ & &
-\ln(x)\ln(1-x) \Li_2(1-x)
\Biggr]
\\
x \otimes \ln(x)\Li_2(1-x)
&=& 
2x \Sf(1-x)  +[1-x +\ln(x)] \Li_2(1-x) \nonumber\\ & &
- \frac{x}{3}
\ln^3(x) - \frac{x}{2} \ln^2(x)
\\
x \otimes x \ln(x)\Li_2(1-x) 
&=& -x \left[\frac{1}{2} \ln^2(x) \Li_2(1-x) - 3 {\rm S}_{1,3}(1-x)
\right]
\\
x \otimes \Sf(1-x)
&=& (1-x) \Sf(1-x) + \frac{x}{6} \ln^3(x)
\\
x \otimes x \Sf(1-x)
&=& x \left[- \ln(x) \Sf(1-x) - 3 \Sa_{1,3}(1-x) \right]
\\
x \otimes \Li_3(x)
&=& \Li_3(x)+\Li_2(x) - x[\zeta(3)+\zeta(2)] - x \ln(x) \nonumber\\ & &
- (1-x) \ln(1-x)
\\
x \otimes x \Li_3(x)
&=& x \left[\zeta(4)-\Li_4(x)\right]
\\
x \otimes \ln(x)\Li_2(x)
&=&  -x \Li_2(1-x) + [1+\ln(x)] \Li_2(x) - x \zeta(2) \nonumber\\ & &
- \ln(1-x)[\ln(x)+2] \nonumber\\ & &
-2x\left[\ln(x) 
- \ln(1-x) +\frac{1}{4}\ln^2(x)\right]
\\
x \otimes x \ln(x)\Li_2(x)
&=& x\left[\Li_4(x) - \ln(x) \Li_3(x) - \zeta(4)\right]
\\
x \otimes \ln^3(1-x)
&=& 6x[\Sf(x) - \zeta(3)] +(1-x) \ln^3(1-x)
\\
x \otimes x \ln^3(1-x)
&=& -6x \left[\zeta(4) - \Sa_{1,3}(x)\right]
\\
x \otimes \ln(x)\ln^2(1-x)
&=& [1-x+\ln(x)] \ln^2(1-x) + 2x[\zeta(2)-\Li_2(x)]\nonumber\\& &
-2x\Biggl[\zeta(3) - \Li_3(x) + \ln(x) \Li_2(x) + \Li_3(1-x)
\nonumber\\ & &
- \ln(1-x) \Li_2(1-x) \Biggr]
\\
x \otimes x \ln(x)\ln^2(1-x)
&=& -2x \left[\frac{1}{4} \zeta(4) 
-{\rm S}_{2,2}(x) + \ln(x) \Sf(x)\right]
\\
x \otimes \ln^2(x)\ln(1-x)
&=& 2x \left[\Li_2(1-x)-\Sf(1-x)\right] \nonumber\\ & &
+ \left[2[1-x + \ln(x)] +\ln^2(x)
\right]
\ln(1-x) \nonumber\\
& & + x \ln(x) \left[2 +\ln(x) +\frac{1}{3} \ln^2(x)\right]
\\
x \otimes x \ln^2(x)\ln(1-x)
&=& - 2 x [ \Sa_{1,3}(1-x) + \frac{1}{6}\ln^3(x)\ln(1-x)]
\end{eqnarray}

For some applications one wishes to give explicit account for soft gluon 
or soft photon resummation using $x$--space formulae. Here one has to 
perform convolutions of the type 
\begin{equation}
\left[\frac{\ln^k(1-x)}{1-x}\right]_+ \otimes 
\left[\frac{\ln^l(1-x)}{1-x}\right]_+ = D_k(x) \otimes D_l(x),~~~k,l \geq 
0~.
\end{equation}
The  Mellin convolution of two $(...)_+$--distributions is again a
$(...)_+$--distribution.
The convolutions $D_k \otimes D_0, k~\epsilon~[0,3]$ are given in 
Eqs.~(69,72,104) and (131) above and can easily be extended to general 
values of $k$ and $l$ using relations given in 
Ref.~\cite{Blumlein:1999if}. The  Mellin transform of $D_k(x)$ reads 
\begin{equation}
\label{eqDD}
\MV[D_k(x)](N) = (-1)^k (k-1)! 
S_{\underbrace{\mbox{\scriptsize 1, \ldots ,1}}_{\mbox{\scriptsize
$k$}}}(N-1)~.
\end{equation}
The harmonic sum in (\ref{eqDD}) obeys a determinant--representation, 
Eq.~(158)~\cite{Blumlein:1999if}, and is a polynomial of only single
harmonic sums. One obtains the Mellin transform of $[\ln(x) 
\ln^{(k)}(1-x)/(1-x)]_+$ by the relation
\begin{equation}
\label{eqDD1}
\MV\left\{\left[\ln(x) \frac{\ln^k(1-x)}{1-x}\right]_+\right\}(N) =
\frac{\partial}{\partial N} 
(-1)^k (k-1)!
S_{\underbrace{\mbox{\scriptsize 1, \ldots ,1}}_{\mbox{\scriptsize
$k$}}}(N-1)~.
\end{equation}
Since the single harmonic sums can be represented by Euler's
$\psi$--function and its derivatives the differentiation for $N$ is
performed easily. Furthermore, one may re-express the result again in a
polynomial of single harmonic sums.
\section{Appendix B:~Mellin Transforms}

\vspace{1mm}
\noindent
Some of the convolutions calculated in the previous section were 
determined using harmonic sums and Mellin transformations of specific 
functions, such as Nielsen--integrals. The Mellin transform of a 
convolution is given by
\begin{equation}
\M\left\{[A \otimes B](x)\right\}(N) = 
\M\left\{[A](x)\right\}(N) \cdot
\M\left\{[B](x)\right\}(N)~.
\end{equation}
Below we list Mellin transforms which were used and were not contained in
Ref.~\cite{Blumlein:1999if}.
\begin{eqnarray}
\M[\ln^p(1-x)](N)   &=& \frac{1}{N} (-1)^p p!
S_{\underbrace{\mbox{\scriptsize 1, \ldots ,1}}_{\mbox{\scriptsize $p$}}}
(N)
\\
\M[\ln^4(1-x)](N)   &=& \frac{1}{N}
\left[S_1^4(N)+6 S_1^2(N) S_2(N) + 3 S_2^2(N) + 8 S1(N) S_3(N) \right.
\nonumber\\ & &  \left.
+6 S_4(N)\right] \nonumber\\
\\
\M[\Sa_{1,p}(x)](N)   &=& \frac{1}{N} \left\{\zeta(p) - \frac{1}
{N}
S_{\underbrace{\mbox{\scriptsize 1, \ldots ,1}}_{\mbox{\scriptsize $p$}}}
(N)
\right\}
\\
\M[\Sa_{1,p}(1-x)](N) &=& - \frac{1}{N} \left[S_p(N) - \zeta(p)\right]
\\
\M[\Sa_{1,3}(x)](N)   &=& \frac{1}{N}\left\{\zeta(4) - \frac{1}{6 N}
\left[S_1^3(N)+3S_1(N)S_2(N)+2S_3(N)\right]\right\}
\\
\M[\Sa_{1,3}(1-x)](N) &=& - \frac{1}{N} \left[S_4(N) - \zeta(4)\right]
\\
\M[\Sa_{1,4}(x)](N)   &=& \frac{1}{N}\Biggl\{\zeta(5) - \frac{1}{24 N}
\Biggl[S_1^4(N)+6S_1^2(N) S_2(N)+8S_1(N)S_3(N) \nonumber\\ & &~~~~
+3S_2^2(N)
    +6 S_4(N)
\Biggr]\Biggr\}
\\
\M[\Sa_{1,4}(1-x)](N) &=& - \frac{1}{N} \left[S_5(N) - \zeta(5)\right]
\\
\M[\Sa_{2,2}(x)](N)   
&=& \frac{\zeta(4)}{4N} - \frac{\zeta(3)}{N^2} 
+ \frac{S_1^2(N)+S_2(N)}{2N^3}
\\
\M[\Sa_{2,2}(1-x)](N)   &=& \frac{1}{N}\left[\frac{1}{4} \zeta(4)
+ S_{1,3}(N) - \zeta(3) S_1(N)\right]
\\
\M\left[\frac{\Sa_{2,2}(1-x)}{1-x}\right](N)   
&=& 2 \zeta(5) - \zeta(2) \zeta(3) - \frac{1}{4} \zeta(4)S_1(N-1)
-S_{1,1,3}(N-1) \nonumber\\ & &
+ \zeta(3) S_{1,1}(N-1)
\\
\M\left[\Li_2^2(x)\right](N) &=&
\frac{1}{N} \Biggl\{\frac{5}{2} \zeta(4)-4\frac{\zeta(3)}{N}
-2\zeta(2) \frac{S_1(N)}{N}
+2 \frac{S_{2,1}(N)}{N}\Biggr\} \nonumber\\ & & +\frac{2}{N^3} \biggl[
S_1^2(N)+S_2(N)\biggr]
\\
\M\left[\ln(1-x) \Li_3(x)\right](N) &=& \frac{1}{N} \Biggl[
-\frac{1}{2} \zeta(2)^2 + \frac{2\zeta(3)}{N} + \zeta(2) \frac{S_1(N)}{N}
- \frac{S_1^2(N)+S_2(N)}{N^2} \nonumber\\ & & - \frac{S_{2,1}(N)}{N}
- \zeta(3) S_1(N) + \zeta(2) S_2(N) - S_{3,1}(N) \Biggr]
\\
\M\left[\ln(x) \Li_3(x)\right](N) &=& - \frac{1}{N^2} \Biggl[\zeta(3)
-\frac{3\zeta(2)}{N} + \frac{S_2(N)}{N} +\frac{3 S_1(N)}{N^2} \Biggr]
\\
\M\left[\ln(x) \Sf(1-x)\right](N) &=& \frac{1}{N}\Biggl\{
3 S_4(N) - 3 \zeta(4) + \frac{1}{N} \left[S_3(N) - \zeta(3) \right]
\Biggr\}
\\
\M\left[\ln(x) \Sf(x)\right](N) &=& \frac{1}{N^2}\Biggl\{
S_1(N) S_2(N) + S_3(N) -  \zeta(2) S_1(N) - 2\zeta(3) \nonumber\\ & &
~~~~+ \frac{1}{N}
\left[S_1^2(N)+S_2(N)\right]\Biggr\}
\\
\M\left[\Li_4(x)\right](N) &=& \frac{\zeta(4)}{N}
-\frac{\zeta(3)}{N^2}+\frac{\zeta(2)}{N^3} - \frac{S_1(N)}{N^4}
\\
\M\left[\ln^2(x) \Li_2(1-x)\right](N) &=& \frac{1}{N}
\Biggl\{6[\zeta(4)-S_4(N)] + \frac{4}{N}[\zeta(3)-S_3(N)]
\nonumber\\ & &
+\frac{2}{N^2}[\zeta(2)-S_2(N)]\Biggr\}\\
\M\left[\frac{\Sa_{2,2}(x)-\zeta(4)/4}{1-x}\right](N) &=& 
-S_{3,1,1}(N-1)+\zeta(3)S_2(N-1)-\frac{\zeta(5)}{2}\\
\M\left[\frac{\Sa_{1,3}(x)-\zeta(4)}{1-x}\right](N) &=& 
S_{2,1,1,1}(N-1)-4\zeta(5)\\
\M\left[\frac{\Sa_{1,3}(1-x)}{1-x}\right](N) &=& 
S_{1,4}(N-1)-\zeta(4)S_1(N-1)+2\zeta(5)-\zeta(2)\zeta(3)
\end{eqnarray}
\section{Appendix C:~Representation of Sums              }

\vspace{1mm}
\noindent
To obtain the inverse Mellin transforms of more complicated convolutions 
back in $x$--space products of harmonic sums and powers of $1/N$ have to 
be expressed as Mellin transforms. A series of representations has been
given in~\cite{Blumlein:1999if} before. Here we list additional 
representations which were used in the present calculation.
\begin{eqnarray}
\frac{S_1(N)}{N} &=& - \M[\ln(1-x)](N)
\\
\frac{S_1(N)}{N^2} &=& - \M[\Li_2(x) - \zeta(2)](N)
\\
\frac{S_1(N)}{N^3} &=&  \M[\Li_3(x) - \zeta(2)\ln(x) - \zeta(3)](N)
\\
\frac{S_1(N)}{N^4} &=& -\M\left[\Li_4(x) - \frac{\zeta(2)}{2}
\ln^2(x) - \zeta(3) \ln(x) - \zeta(4)\right](N)
\\
\frac{S_2(N)}{N} &=& -\M[\Li_2(1-x) - \zeta(2)](N)
\\
\frac{S_2(N)}{N^2} &=&  \M[2 \Sf(1-x) + \ln(x)[\Li_2(1-x) -
\zeta(2)]](N)
\\
\frac{S_2(N)}{N^3} &=& 
-\M\left[3 \Sa_{1,3}(1-x) + 2 \ln(x) \Sf(1-x)
+\frac{1}{2}\ln^2(x)[\Li_2(1-x) - \zeta(2)]\right](N)
\nonumber\\
\\
\frac{S_3(N)}{N} &=& -\M[\Sa_{1,2}(1-x) - \zeta(3)](N)
\\
\frac{S_3(N)}{N^2} &=& \M[3\Sa_{1,3}(1-x)+\ln(x) \Sf(1-x) - \zeta(3)
\ln(x)](N)
\\
\frac{S_4(N)}{N} &=& -\M[\Sa_{1,3}(1-x) - \zeta(4)](N)
\\
\frac{S_1^2(N)}{N} &=& \M[\ln^2(1-x)+\Li_2(1-x) -\zeta(2)](N)
\\
\frac{S_1^2(N)}{N^2} &=& 
-\M[2\Sf(x)+2\Sf(1-x)+\ln(x)[\Li_2(1-x)-\zeta(2)] -2\zeta(3)](N)
\\
\frac{S_1^2(N)}{N^3} &=& \M\left[2\Sa_{2,2}(x) - \frac{\zeta(4)}{2}
- 2 \zeta(3) \ln(x)\right](N)\nonumber\\ & &
+\M\left[3 \Sa_{1,3}(1-x) + 2 \ln(x) \Sf(1-x)
+\frac{1}{2}\ln^2(x)
[\Li_2(1-x) - \zeta(2)]\right](N) \nonumber\\
\\
\frac{S_2^2(N)}{N} &=& \M\left[4 \Sa_{22}(x) + \Sa_{13}(1-x)
        - 4 \ln(x)\Sf(x) - 2 \ln(x)\ln(1-x)\Li_2(x)\right]
\nonumber\\
 & &    + \M\left[ 4 \ln(1-x)\Li_3(x) + \Li_2^2(x) 
        - 4\zeta(3)\ln(1-x) - \zeta(4) \right]
\\
\frac{S_2(N) S_1(N)}{N} &=& 
\M\left[\Sf(1-x)-\Li_3(1-x) + \ln(1-x)[\Li_2(1-x)
- \zeta(2)]\right](N)
\\
\frac{S_2(N) S_1(N)}{N^2} &=& \M\left[\ln(x)[\Sf(x)-\Sf(1-x)]
-3\Sa_{1,3}(1-x)\right](N) \nonumber\\ & &
- \M\left[2\Sa_{2,2}(x) + \zeta(2) \Li_2(x) - \zeta(3) \ln(x)
- 3\zeta(4)\right](N)
\\
\frac{S_3(N) S_1(N)}{N} &=&  \M\left[\Sa_{2,2}(1-x)+\Sa_{1,3}(1-x)
- \frac{1}{2} \Li_2^2(1-x) \right](N) \nonumber\\ & &
\nonumber\\
& &
+ \M\left[[\ln(1-x)[\Sf(1-x) -\zeta(3)]\right](N)
\\
\frac{S_2(N) S_1^2(N)}{N} &=& \M\left [- 4 \Sa_{22}(x) - 2 \Sa_{22}(1-x)
    + 2 \Sa_{13}(x) - \Sa_{13}(1-x) + 4 \ln(x)\Sf(x) \right]
\nonumber\\
 & & 
    + \M\left[ - 2\ln(1-x)\Sf(1-x) - 4 \ln(1-x)\Li_3(x) 
    + \Li_2^2(1-x) - \Li_2^2(x) \right]
\nonumber\\
 & &
    + \M\left[ 2 \ln(x)\ln(1-x)\Li_2(x)
    + \frac{1}{3}\ln(x)\ln^3(1-x) + \zeta(2)\ln^2(1-x) \right]
\nonumber\\
 & &
    + \M\left[ 4\zeta(3)\ln(1-x) - \zeta(4) \right]
\end{eqnarray}

\vspace{1mm}\noindent
{\bf Acknowledgment.} This paper was supported in part by
DFG Sonderforschungsbereich Transregio~9, Computergest\"utzte Theoretische
Physik.
\newpage
\begin{center}

\mbox{\epsfig{file=fig1.eps,height=9cm,width=10cm}}

\vspace{2mm}
\noindent
\small
\end{center}
{\sf
Figure~1:~
The non--singlet radiation function $D_{\rm NS}(x,Q^2)$ to $O((\alpha 
L)^5)$ as a function of $x$ for $Q= 10 \GeV, 100 \GeV$ and $1 \TeV$.}
\normalsize

\vspace*{1cm}
\begin{center}

\mbox{\epsfig{file=fig2.eps,height=9cm,width=10cm}}

\vspace{2mm}
\noindent     
\small   
\end{center}
{\sf
Figure~2:~
The effect of the resummed contributions beyond $O((\alpha L)^5)$ compared 
to the contributions up to $O((\alpha L)^5)$ for $D_{\rm NS}(x,Q^2)$ as a 
function of $x$ and $Q$.}
\normalsize

\vspace*{1cm}
\newpage
\begin{center}

\mbox{\epsfig{file=fig3.eps,height=9cm,width=9cm}}

\vspace{2mm}
\noindent     
\small   
\end{center}
{\sf
Figure~3:~
Relative contribution of the first order non-singlet radiator compared to 
all terms up to $O((\alpha L)^5)$ as a function of $x$ and $Q$ and relative 
correction of all contributions up to $O((\alpha L)^4)$ in comparison to 
the terms up to $O((\alpha L)^5)$.}
\normalsize

\vspace*{1cm}
\begin{center}

\mbox{\epsfig{file=fig4.eps,height=9cm,width=10cm}}

\vspace{2mm}
\noindent     
\small   
\end{center}
{\sf
Figure~4:~
The polarized singlet contributions $D_{ij}$ as a function of $x$ and $Q$ 
in $\%$. Here $D_{11}$ denotes the pure singlet term, which has to be 
added 
to the non--singlet contribution.} 
\normalsize
\newpage
\begin{center}

\mbox{\epsfig{file=fig5.eps,height=9cm,width=9cm}}

\vspace{2mm}
\noindent     
\small   
\end{center}
{\sf
Figure~5:~Relative contribution of the first order singlet radiators 
$D_{ij}^1$ in the terms to $O((\alpha L)^5)$.}
\normalsize

\vspace*{1cm}
\begin{center}

\mbox{\epsfig{file=fig6.eps,height=9cm,width=9cm}}

\vspace{2mm}
\noindent     
\small   
\end{center}
{\sf
Figure~6:~Relative contribution of all terms of the singlet radiators 
$D_{ij}$ to the 4th order in $\alpha 
L$ if compared to all terms to 5th order as a function of $x$ and $Q$.}
\normalsize

\newpage
\begin{center}

\mbox{\epsfig{file=fig7.eps,height=9cm,width=10cm}}

\vspace{2mm}
\noindent     
\small   
\end{center}
{\sf
Figure~7:~
The polarized singlet contributions $D_{ij}(x,Q^2)$ corresponding to the 
resummation of the $O(\alpha \ln^2(x))$ terms as a function of $x$ and $Q$ 
in $\%$ starting with $O(\alpha^2)$.  
Here $D_{11}$ contains also the non--singlet contribution.}
\normalsize

\vspace{5mm}
\noindent
\begin{minipage}[b]{.46\linewidth}
\centering\epsfig{figure=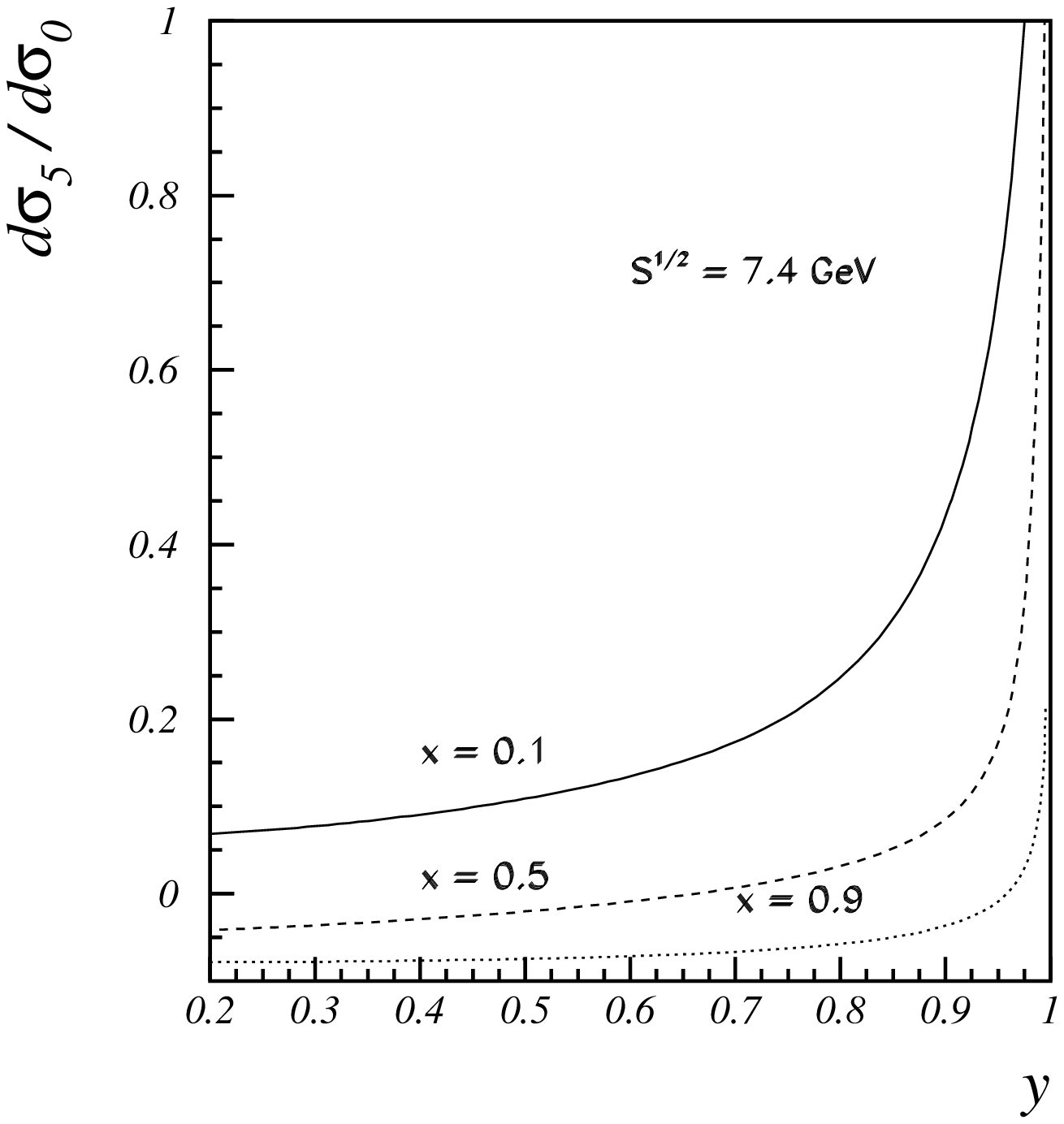,width=\linewidth}
\end{minipage}\hfill
\begin{minipage}[b]{.46\linewidth}
\centering\epsfig{figure=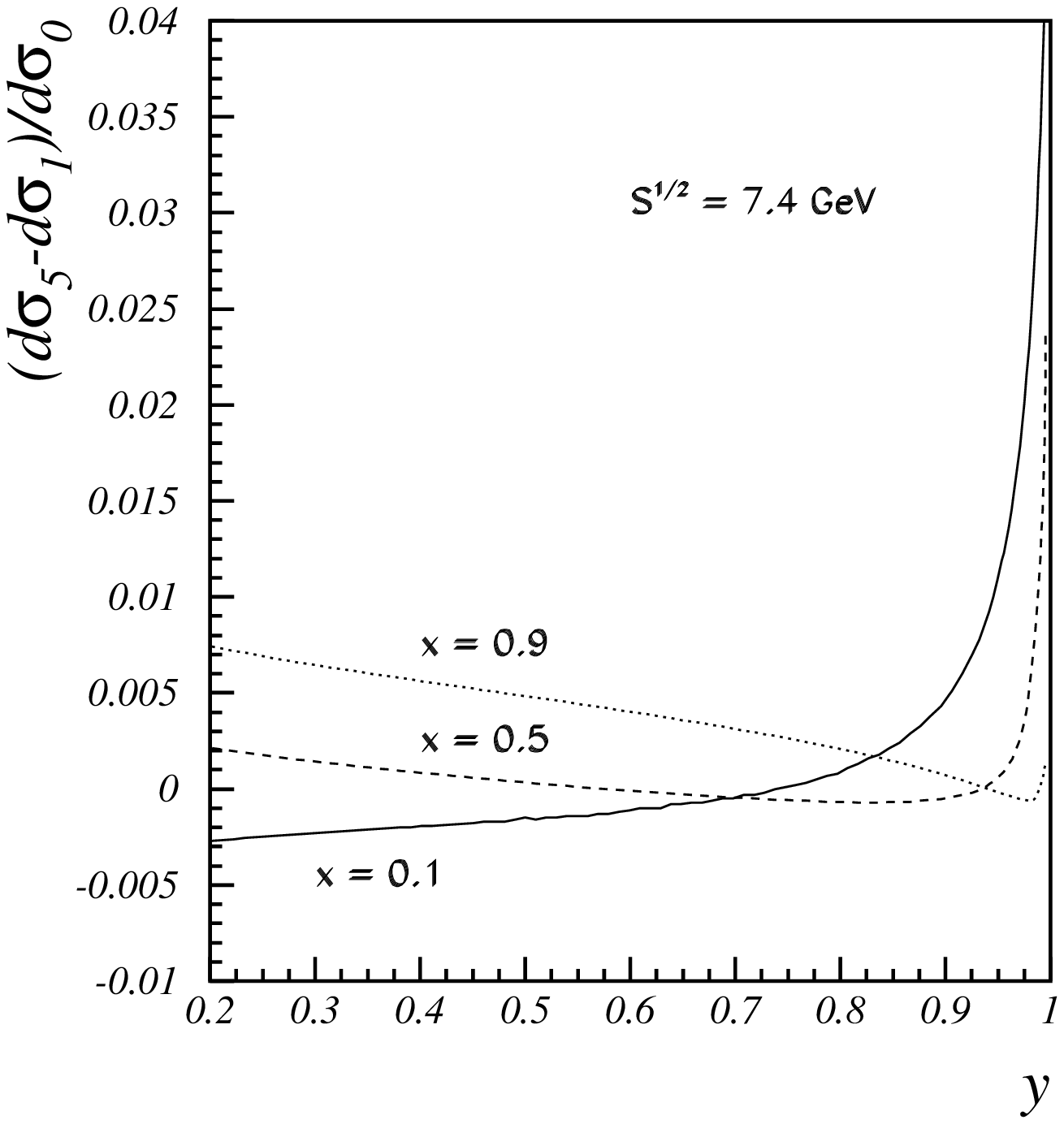,width=\linewidth}
\end{minipage}   

\vspace{2mm}\noindent
%
{\sf
Figure~8:~
The QED radiative initial state correction to the differential 
cross 
section of polarized deep--inelastic lepton--proton scattering including 
the $O((\alpha L)^5)$ corrections as a function of $x$ and $y$ in a fixed 
target experiment $(E_e = 27.5 \GeV)$.
Left~: the correction factor. 
Right~: the contributions due to non--leading orders.}
\normalsize

\vspace{5mm}
\noindent
\begin{minipage}[b]{.46\linewidth}
\centering\epsfig{figure=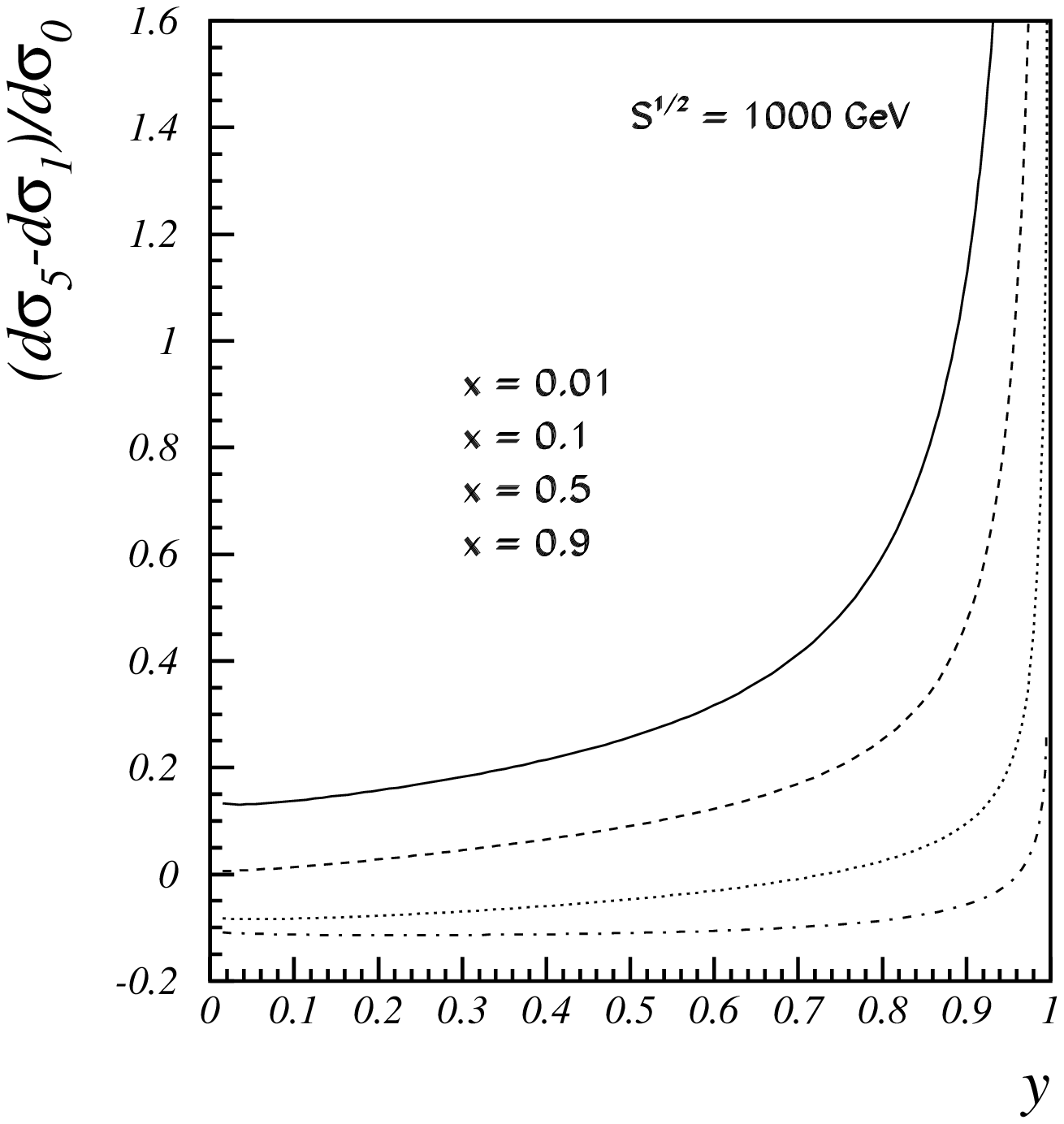,width=\linewidth}
\end{minipage}\hfill
\begin{minipage}[b]{.46\linewidth}
\centering\epsfig{figure=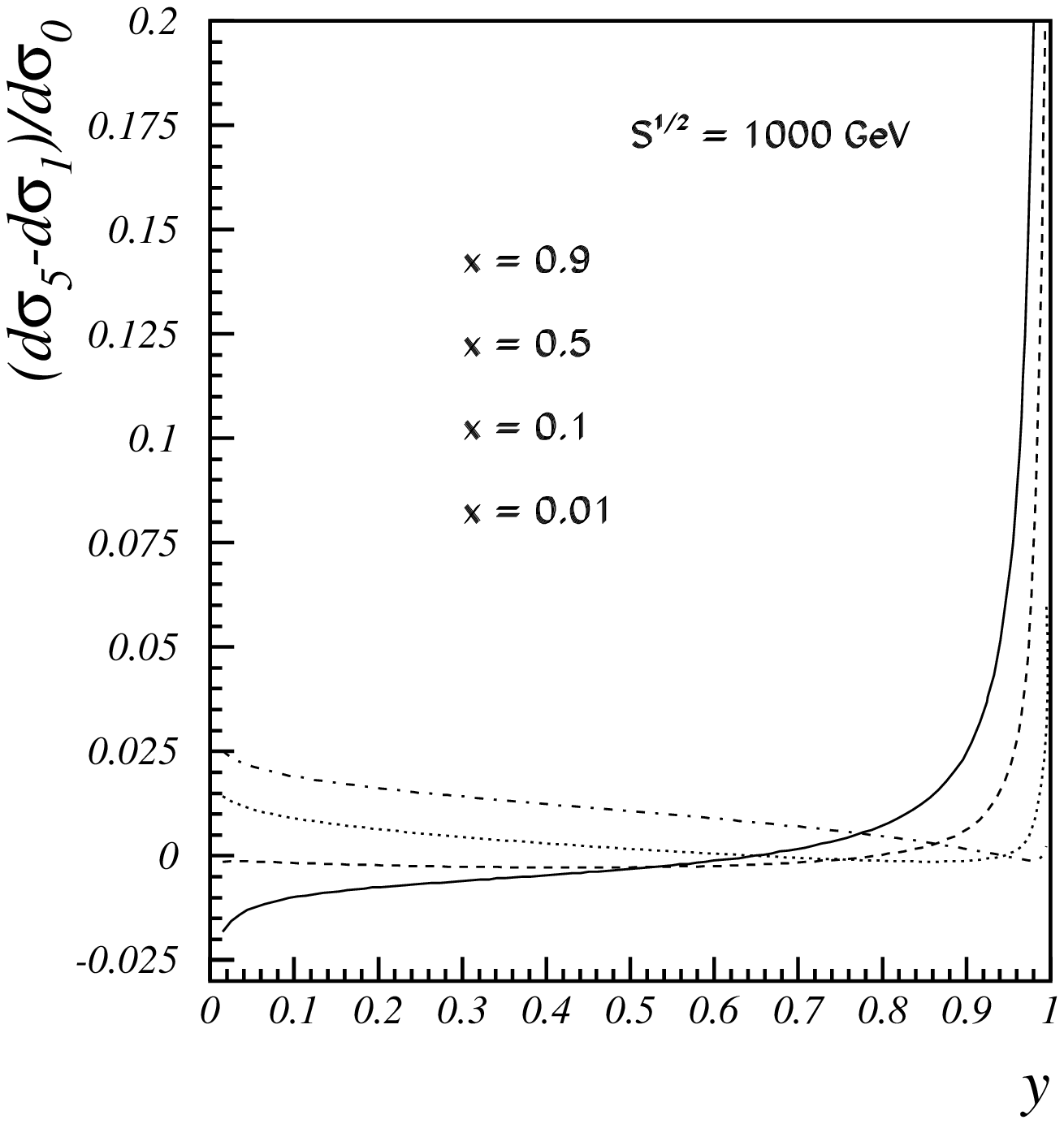,width=\linewidth}
\end{minipage}   

\vspace{2mm}\noindent
%
{\sf
Figure~9:~
The QED radiative initial state correction to the differential cross
section of polarized deep--inelastic lepton--proton scattering including
the $O((\alpha L)^5)$ corrections as a function of $x$ and $y$ for a 
collider experiment at $S = 1 \TeV^2$. 
Left~: the correction factor. 
Right~: the contributions due to non--leading orders.}
\normalsize


\end{document}